\documentclass[
    aps,
    prl,
    amsmath,amssymb,
    tightenlines, 
    superscriptaddress,
    floatfix,
    reprint,
]{revtex4-2}

\usepackage[american]{babel} 
\usepackage[utf8x]{inputenc} 

\usepackage{newtxtext,newtxmath} 
\usepackage{microtype} 

\usepackage{bm}
\usepackage{mathtools} 
\usepackage{physics} 
\usepackage{dsfont} 
\usepackage{braket} 
\usepackage{stackengine} 
\usepackage{lipsum} 
\usepackage{amsmath} 

\usepackage[cal=cm,scr=dutchcal]{mathalpha} 

\usepackage{graphicx}
\usepackage[svgnames]{xcolor}
\usepackage{siunitx} 

\usepackage{ragged2e} 
\usepackage{array} 
\usepackage{tabularx} 
\usepackage{booktabs} 
\usepackage[normalem]{ulem} 

\definecolor{cset-aps-blueberry}{RGB}{28,128,158}
\definecolor{cset-aps-blue}{RGB}{46,44,184}
\definecolor{cset-aps-turquoise}{RGB}{0,67,88}
\definecolor{cset-aps-limegreen}{RGB}{190,219,67}
\definecolor{cset-aps-green}{RGB}{31,138,112}
\definecolor{cset-aps-yellow}{RGB}{255,225,25}
\definecolor{cset-aps-orange}{RGB}{253,116,0}
\definecolor{cset-aps-red}{RGB}{219,0,43}

\definecolor{cset-aps-kobalt-medium}{RGB}{62,54,222}
\definecolor{cset-aps-kobalt-dark}{RGB}{28,24,150}

\definecolor{cset-aps-my-label-red}{RGB}{202,0,17}
\definecolor{cset-aps-my-label-blue}{RGB}{53,71,140}
\definecolor{cset-aps-my-label-gray}{RGB}{145,145,145}

\usepackage{hyperref}
\hypersetup{%
    colorlinks=true,
    linkcolor={cset-aps-blue},
    linkbordercolor={cset-aps-red},
    filecolor={cset-aps-orange},
    filebordercolor={cset-aps-orange},
    citecolor={cset-aps-kobalt-medium},    
    citebordercolor={cset-aps-kobalt-medium},
    urlcolor={cset-aps-kobalt-dark},
    urlbordercolor={cset-aps-kobalt-dark},
    menucolor={cset-aps-limegreen},
    menubordercolor={cset-aps-limegreen},
    breaklinks=true,
    pdfborderstyle={/S/U/W 2},
    pdfpagemode=UseOutlines,
    pdfstartpage={1}
}
\usepackage{enumitem} 

\newcommand{\ii}{\text{i}}

\newcommand{\rabi}{\Omega_\text{R}}

\newcommand*{\figref}[2][]{%
	Fig.~\hyperref[{#2}]{%
		\ref*{#2}%
		\ifx\\#1\\%
		\else
		(#1)%
		\fi
	}%
}

\newcommand*{\figureref}[2][]{%
	Figure~\hyperref[{#2}]{%
		\ref*{#2}%
		\ifx\\#1\\%
		\else
		(#1)%
		\fi
	}%
}

\newcommand{\orcid}[1]{\href{https://orcid.org/#1}{\includegraphics[width=7pt]{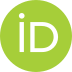}}}
\newcommand{\affTUDa}{Technische Universit{\"a}t Darmstadt, Fachbereich Physik, Institut f{\"u}r Angewandte Physik, Schlossgartenstra{\ss}e 7, 64289 Darmstadt, Germany}
\newcommand{\affHFHF}{Helmholtz Forschungsakademie Hessen f{\"u}r FAIR (HFHF), Campus Darmstadt, Schlossgartenstra{\ss}e 2, 64289 Darmstadt, Germany}

\allowdisplaybreaks 

\AtBeginDocument{%
}

\begin{document}

\title[Dichroic Mirror Pulses for Bragg Diffraction]{
Dichroic Mirror Pulses for Optimized Higher-Order Atomic Bragg Diffraction
}

\author{D. Pfeiffer\,\orcid{0000-0003-4530-3251}}
\email{apqpub@physik.tu-darmstadt.de}
\affiliation{\affTUDa}
\author{M. Dietrich\,\orcid{0000-0003-1526-6293}}
\homepage[]{https://www.iap.tu-darmstadt.de/tqo}
\affiliation{\affTUDa}
\author{P. Schach\,\orcid{0000-0002-6672-9692}}
\homepage[]{https://www.iap.tu-darmstadt.de/tqo}
\affiliation{\affTUDa}
\author{G. Birkl\,\orcid{0000-0002-4137-9227}}
\homepage[]{https://www.iap.tu-darmstadt.de/apq}
\affiliation{\affTUDa}
\affiliation{\affHFHF}
\author{E. Giese\,\orcid{0000-0002-1126-6352}}
\homepage[]{https://www.iap.tu-darmstadt.de/tqo}
\affiliation{\affTUDa}



\begin{abstract}
Increasing the sensitivity of light-pulse atom interferometers progressively relies on large-momentum transfer techniques. 
Precise control of such methods is imperative to exploit the full capabilities of these quantum sensors. 
One key element is the mitigation of deleterious effects such as parasitic paths deteriorating the interferometric signal. 
In this Letter, we present the experimental realization of dichroic mirror pulses for atom interferometry, its scalability to higher-order Bragg diffraction, and its robustness against initial momentum spread. 
Our approach selectively reflects resonant atom paths into the detected interferometer output, ensuring that these contribute to the signal with intent. 
Simultaneously, parasitic paths are efficiently transmitted by the mirror and not directed to the relevant interferometer outputs. 
This method effectively isolates the desired interferometric signal from noise induced by unwanted paths.
It can be readily applied to existing setups capable of higher-order Bragg diffraction.\\
\\
\noindent
\textcolor{black}{
This article has been published in \href{https://doi.org/10.1103/PhysRevResearch.7.L012028}{Physical Review Research \textbf{7}, L012028 (2025)} under the terms of the \href{https://creativecommons.org/licenses/by/4.0/}{Creative Commons Attribution License 4.0 [CC BY]}}. DOI: \url{https://doi.org/10.1103/PhysRevResearch.7.L012028}
\end{abstract} 

\maketitle
Large momentum transfer (LMT)~\cite{debs:2011:cold-atom,chiow:2011:102} is one promising approach for scaling the sensitivity of light-pulse atom interferometers.
By complementing architectures based on extending free-fall times in microgravity~\cite{muentinga:2013:interferometry,Barrett:2016:dual,Lachmann:2021:ultracold,williams:2024:pathfinder}, large-scale fountain setups~\cite{schlippert:2020:matterwave,Badurina:2020:aion,Zhou:2011:development,kovachy:2015:quantum}, and relaunch geometries~\cite{impens:2009:spacetime,Schubert:2021:multi-loop,schubert:2019:scalable,abend:2016:atom,hughes:2009:suspension}, LMT promotes atom interferometers to high-precision quantum sensors. 
Among other achievements, it has facilitated the most precise determination of the fine-structure constant~\cite{morel:2020:determination,parker:2018:measurement}, the generation of record-scale spatial superpositions~\cite{kovachy:2015:quantum}, and competitive tests of relativity~\cite{asenbaum:2020:atom-interferometric}. 
While today's technology enables momentum transfer of hundreds of photon recoils~\cite{gebbe:2021:twin,morel:2020:determination,parker:2018:measurement,Canuel:2020:elgara,zhan:2020:zaiga}, proposals for atom-interferometric gravitational-wave or dark-matter detection~\cite{abend:2024:terrestrial,Badurina:2020:aion,Abe:2021:matterwave} rely on a transfer exceeding $10^2$ to $10^4$ photon recoils and thus on optimizing LMT. 
Among the most prominent LMT techniques are (i) Bloch oscillations~\cite{clade:2009:large,mcDonald:2013:80hbark,pagel:2020:symmetric,rahman:2024:bloch,fitzek:2023:accurate}, 
(ii) sequential pulses~\cite{rudolph:2020:largemomentum,berg:2015:composite-light,mcguirk:2000:largearea}, 
(iii) double diffraction~\cite{leveque:2009:enhancing,giese:2013:double,kueber:2016:experimental}, 
and (iv) higher-order diffraction~\cite{Abe:2021:matterwave,hartmann:2020:regimes,hartmann:2020:atomicraman,mueller:2008:atom}
with many experiments combining several techniques~\cite{abend:2016:atom,mueller:2008:atom,morel:2020:determination,parker:2018:measurement,ahlers:2016:doublebragg, mueller:2009:atom}. 

Although each method has its own characteristics, all are susceptible to imperfect momentum transfer and pulse infidelity~\cite{chiarotti:2022:prx}, so parasitic momenta are populated inherently. 
These imply overall loss, diffraction phases~\cite{buechner:2003:diffraction,estey:2015:high-resolution}, and introduce unintended paths to the interferometer. 
These paths are redirected into the detected output ports of the interferometer, causing multipath interference~\cite{beguin:2022:characterization,Altin:2013:precision,parker:2016:controlling,jenewein:2022:bragg} deteriorating the sensitivity. 
Current mitigation techniques include (i) tailored design of pulse shapes to suppress the population of parasitic momenta ~\cite{wilkason:2022:atominterferometry,rodzinka:2024:optimalfloquetengineeringlarge}, sometimes leveraged by optimal control~\cite{chen:2023:enhancing,Saywell:2020:optimalcontrol,saywell:2020:biselective,li:2024:robust,louie:2023:robust,chiarotti:2022:prx} and applied to the full pulse sequence~\cite{saywell:2023:enhancing}, (ii) optimal-control techniques to enhance the signal of multipath interference~\cite{wang:2024:arxiv}, or (iii) exploitation of destructive interference of parasitic paths~\cite{beguin:2023:atominterferometry}. 

We present a demonstration of a complementing technique optimizing the mirror pulses of the interferometer.
Our approach has been inspired by a proposal~\cite{kirsten:2023:prl,kirsten:2023:phd} of momentum-selective, i.\,e., dichroic mirror pulses (DMP) based on Bragg diffraction~\cite{siemss:2020:analytic,giese:2015:mechanisms} that only redirect the two intentionally populated, i.\,e., resonant paths (distinguishable through their momenta), while being made transparent for the dominant parasitic paths, which are not redirected to the detected output of the interferometer. 
Such an evolution is induced by applying a pulse area of $\pi$ to the resonant paths, while parasitic orders experience pulse areas of multiples of $2\pi$.
We establish the scalability of this technique through an examination of resonant third- and fifth-order diffraction.
Accompanying simulations prove that the method exhibits robust performance across a wide range of momentum spreads instead of being velocity selective.
As a consequence, this technique can be readily applied to existing setups capable of higher-order Bragg diffraction.

\begin{figure}[t!]
    \centering
    \includegraphics[width=1\linewidth]{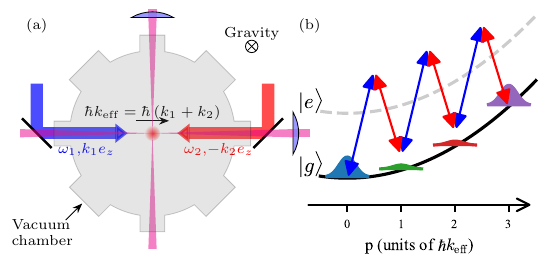}
    \caption{
    Experimental setup and energy band structure of higher-order Bragg diffraction.
    (a) An atom ensemble with narrow momentum distribution is prepared via Bose-Einstein condensation in a crossed dipole trap (transparent red).
    Two counterpropagating laser beams (blue and red) 
    induce Bragg diffraction and transfer momenta $n \hbar k_\text{eff}$ as beam splitters and mirrors.
    (b) Energy-momentum conservation in a third-order Bragg process.
    Blue and red arrows denote the induced absorption and emission of photons of the corresponding Bragg beams.
    The light is strongly detuned with respect to one-photon transitions.
    }
    \label{fig:setup}
\end{figure}
To induce Bragg diffraction, two counterpropagating laser beams with frequency difference $\Delta \omega=\omega_1-\omega_2$ couple two momenta via a virtual state, see \figref{fig:setup}. 
Subsequently absorbing a photon from one laser field and re-emitting it into the counterpropagating field transfers momentum $\hbar k_\text{eff} = \hbar (k_1 + k_2)$ to the atom as a consequence of momentum conservation, where $ k_j$ is the modulus of the wave vector of field $j$. 
For frequencies $\omega_{1,2}$ far detuned from the one-photon transition, Rabi oscillations between the two resonant momenta are induced at a two-photon Rabi frequency $\rabi$. 
Coupling the initial momentum $p_0$ to higher momentum states is possible by transferring $n \hbar k_\text{eff}$ momenta through a $2n$-photon process~\cite{shore:1990:theory}, resonantly populating the momentum $p_n= p_0 + n \hbar  k_\text{eff} $. 
For an atom of mass $m$, the resonance condition is given by $\Delta \omega =n \hbar k_\text{eff}^2 / (2m)  + p_0k_\text{eff}/m$ [see \figref[b]{fig:setup}], where the effective $2n$-photon Rabi frequency scales with the $n$th power of $\rabi$. 
In this Letter, we focus on third- and fifth-order Bragg diffraction $n=3$ and $n=5$ which demonstrate all relevant mechanisms, especially since the dominating parasitic momenta $p_i$ are inherently given by $i=1$ and $i=n-1$~\cite{kirsten:2023:prl,kirsten:2023:phd}. 

For the preparation of ultracold ensembles of $^{87}$Rb, we produce in a far-detuned crossed optical dipole trap a Bose-Einstein condensate (BEC) of typically $20000$ atoms (temperature: $\SI{25\pm 5}{\nano\kelvin}$, condensate fraction $\geq 80\,\%$) by forced evaporative cooling~\cite{lauber:2011:pra}. 
The ensemble in the atomic ground state $\ket{g}=\ket{5^2S_{1/2},F=1}$ is released with momentum $p_0=0$ along the direction of the Bragg beams. 
An expansion time of $\SI{3}{\milli\second}$ converts the ensemble's mean field energy to kinetic energy, after which 
the momentum spread of the ensemble (fitted to a Gaussian) is $\Delta p=0.13(3)\hbar k_\text{eff}$. 
To induce the desired order of diffraction, the frequency difference of a pair of counterpropagating horizontally-aligned laser beams is adjusted to the respective resonance condition $\Delta\omega=n\times2\pi\times\SI{15.1}{\kilo\hertz}$. 
The detuning $2\pi\times\SI{3.2}{\giga\hertz}$ to the state manifold $\ket{e}=\ket{5^2P_{3/2},F=2}$ is sufficiently large to suppress its population by one-photon absorption. 
The waist $w_0=\SI{1170\pm 50}{\micro\meter}$ of the Bragg beams significantly exceeds the size of the expanded atom cloud ($\leq\SI{100}{\micro\meter}$) and the distance traveled by the atoms ($\leq\SI{250}{\micro\meter}$). 
We apply a pulse with a Blackman window function $f(t)=0.42-0.5\cos\left(2\pi t/\tau\right)+0.08\cos\left(4\pi t/\tau\right)$ defined for a duration $0 \leq t \leq \tau$ with FWHM $\simeq 0.405 \tau$ and $f(\tau/2)=1$ as smooth pulse envelope. 
After a Bragg-pulse sequence, a time of flight (TOF) of typically $\SI{15}{\milli\second}$ projects the final atomic momentum distribution to the far-field and spatially separates different momenta. 
Detection is performed by resonant-absorption imaging. 
\begin{figure*}[!htb]
    \centering

    \includegraphics[width=1\linewidth]{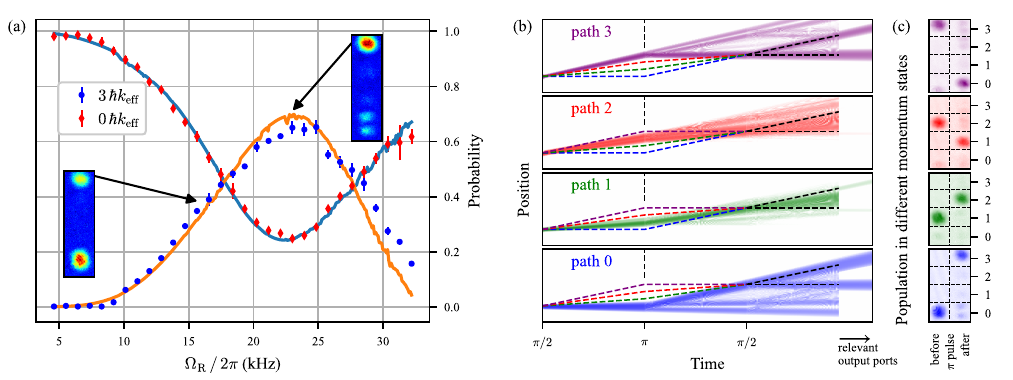}
    \caption{
    Experiment and simulation of third-order Bragg diffraction.
    (a) Resonant Rabi scan for a fixed duration $\tau=\SI{90}{\micro\second}$ of a Blackman pulse.
    Scanning $\rabi$ induces Rabi oscillations between momentum states $p_0$ and $p_3$ (measured probabilities as red diamonds and blue dots).
    Solid lines show a corresponding simulation with a fitted width $\Delta p=0.13(1)\hbar k_\text{eff}$ of a Gaussian momentum distribution.
    False-color insets depict momentum distributions after a beam splitter (left) and mirror pulse (right) with population in $p_0$ at the bottom and in $p_3$ on top. The undeflected density lobes of $p_{0}$ are caused by momentum selectivity.
    (b) The simulated time evolution separated into two resonant ($0,3$) and two parasitic ($1,2$) paths on a logarithmic color map, with the deleted paths sketched by colored dashed lines.
    The dashed black lines after the second beam splitter indicate the relevant output ports.
    In the final section, all population that is not detected in these ports is blanked.
    The simulation demonstrates that also parasitic orders are reflected and coupled into the output ports.
    (c) Confirming these numerical findings, we present experimental absorption images of momentum distributions before (left) and after (right) the mirror pulse.
    The resonant orders (purple and blue) are efficiently reflected, but so are the parasitic orders (red and green), which are redirected to the exit ports with high efficiency. 
    }\label{fig:init_sit}
\end{figure*}

Higher-order diffraction requires longer pulses or increased laser power, implying an increased $\Omega_\text{R}$. 
The two parameters are balanced to minimize velocity selectivity while still observing a two-level behavior between resonant momenta.
However, this quasi-Bragg regime~\cite{mueller:2008:atom-wave,beguin:2022:characterization,siemss:2020:analytic,plotkin:2018:three-path} leads to inevitable population of parasitic orders $p_{i}$. 
By using smooth pulse envelopes (like Blackman pulses) this issue can be mitigated, but not eliminated. 

Fixing $\tau=\SI{90}{\micro\second}$ and scanning $\rabi$ by varying the optical power of the pulse
~\footnote{
We use $\rabi = 0.42 \times 4P \Tilde{U}_0(\lambda)/(\hbar \pi w_0^2)$, where $P$, $w_0$, and $\lambda=\SI{780.226}{\nano\meter}$ are average power, waist, and wavelength of the Bragg beams. $\Tilde{U}_0(\lambda)$ is the dipole factor~\cite{kuber:2014:dynamics,metcalf:1999:springer,grimm:1999:dipole,steck:2007:quantum} for $^{87}$Rb.
The factor $0.42$ arises from $f(t)$.
This definition has been verified by finding good agreement of $\rabi$ experimentally observed for resonant first-order diffraction and the one extracted from simulations.
}
when tuned to third-order resonance $n=3$, we determine the probability of populating momenta $p_{i}$ by extracting the atom numbers in the relevant orders $i=0,1,2,3$ and normalizing to their sum. 
In \figref[a]{fig:init_sit} we show the probabilities of $p_0$ (red diamonds) and $p_3=p_0+3\hbar k_\text{eff}$ (blue dots) as a function of $\rabi$. 
From this scan we extract $\rabi$ for beam splitter ($\pi/2$) and mirror ($\pi$) pulses for third-order diffraction.
The transfer to momentum $p_3$ is limited to $65\,\%$, which we attribute to velocity selectivity \cite{Szigeti:2012:momentumwidth}. 

Our experimental observations are supported by simulations implementing the effective Hamiltonian~\cite{mueller:2008:atom-wave} 
\begin{equation}\label{hamiltonian}
    \hat{H}=\frac{\hat{p}^2}{2m} + 2\hbar\Omega_\text{R} f(t)\cos^2\left(\frac{k_\text{eff}\hat{x}-\Delta \omega t+\phi}{2}\right),
\end{equation}
with the phase of the Bragg beams $\phi$ and $[\hat{x},\hat{p}]=\ii \hbar $, since atomic interactions can be neglected due to the low atom density after expansion.
For these simulations, we use a Split-Step-Fourier method implementing the palindromic \emph{PP 3/4 A scheme}~\cite{auzinger:2017:practical}. 
Since our figure of merit, the diffracted population, is not expected to depend on the laser phase in Bragg diffraction, we choose $\phi=0$ for our simulations.
We fit our numerical model to the data using the experimental values of $\rabi$ and $\tau$ and only leaving the initial spread $\Delta p$ 
as a free parameter. 
The results are shown as solid lines in \figref[a]{fig:init_sit} and agree well with our experiments. 
Applying the same routine to a larger data set varying $\SI{50}{\micro\second}\leq\tau\leq\SI{150}{\micro\second}$, we infer
$\Delta p = 0.13(1)\hbar k_\text{eff}$, matching the experimental value.

Using the parameters obtained from this fit, we extend our simulations to a full Mach-Zehnder ($\pi/2$ - $\pi$ - $\pi/2$) interferometer (MZI). 
Besides the resonant paths $p_0$ and $p_3$, parasitic paths emerge after the first beam splitter. 
To demonstrate their impact, we show in \figref[b]{fig:init_sit} a path-resolved version of the MZI simulation. 
The atom density is displayed on logarithmic scale with the input to every path normalized to unity. 
Dashed colored lines indicate the paths of non-displayed orders and dashed black lines the output ports of the MZI associated with $p_0$ and $p_3$. 
The final section of the time evolution hides the atoms not detected in these ports to augment the contribution of each path to the signal.
The resonant paths 0 and 3 suffer loss from velocity selectivity, but the bulk of the population couples into the output ports as intended. 
Because the mirror pulse also redirects the parasitic paths 1 and 2, they are coupled into the relevant output ports, interfering with resonant paths and corrupting the signal. 
While the effect is small for adiabatic pulses that suppress the initial population of parasitic paths, it can limit the sensitivity of the interferometer~\cite{kirsten:2023:prl}. 
To verify the numerical observation that the mirror pulse redirects both resonant and parasitic paths, we selectively prepare atoms in four input states $p_\text{in} \in \left\{p_0, p_1, p_2, p_3\right\}$ by applying Bragg pulses at the respective resonance, as shown in the left column of \figref[c]{fig:init_sit} by absorption images in the far field. 
After $\SI{4}{\milli\second}$ of propagation, we apply the third-order mirror pulse ($\tau=\SI{90}{\micro\second}$, $\rabi=2\pi\times\SI{23\pm 2}{\kilo\hertz}$). 
The right column of \figref[c]{fig:init_sit} presents the resulting momentum distributions. 
The resonant orders (purple and blue) are efficiently reflected, but so are the parasitic orders (red and green), which are redirected to the exit ports with high efficiency. 

To overcome this problem, we implement a DMP that is reflective only for the resonant paths 0 and 3 while not redirecting the parasitic paths 1 and 2. 
Similar to the experiment presented in \figref[c]{fig:init_sit}, we prepare ensembles in all four relevant $p_\text{in}$
and apply a mirror pulse resonant to third-order diffraction, varying $\tau$ and $\rabi$. 
To obtain the reflectivity $R_{\text{in,out}}$ for each path, we measure the final probability of all four output momenta $ p_\text{out}$ by integrating over the respective momentum distribution in the far field and normalizing it.
In \figref[a]{fig:final_sit} (left column) we show the measured values for $R_{0,3}$ and $R_{1,2}$. 
Corresponding simulations are depicted in the right column, showing excellent agreement. 
The white dashed line indicates the Rabi-frequency scan of \figref[a]{fig:init_sit}. 
Since resonant paths are redirected by six-photon processes but parasitic paths by two-photon transitions, their multi-photon Rabi frequencies differ, as can be seen from different oscillation periods in the top and bottom panels. 
This behavior suggests a parameter set where resonant orders experience a pulse area $\pi$, while parasitic orders experience pulse areas of multiples of $2\pi$. 
In fact, for $\tau=\SI{120}{\micro\second}$ and $\rabi=2\pi\times\SI{21(2)}{\kilo\hertz}$ we observe a DMP with $R_{0,3}= 0.62(1)$ and $R_{1,2} = 0.08(1)$ (white diamonds).
In comparison to the parameters of \figref[a]{fig:init_sit} with $R_{0,3}= 0.65(2)$ and $R_{1,2} = 0.72(1)$ (black crosses), the reflectivity of parasitic paths drops significantly while the one of resonant paths remains almost unaltered.
The pronounced reflectivity of resonant paths combined with the low reflectivity of parasitic paths implements the intended DMP. 
\begin{figure*}[ht!]
    \centering
   \includegraphics[width=1\linewidth]{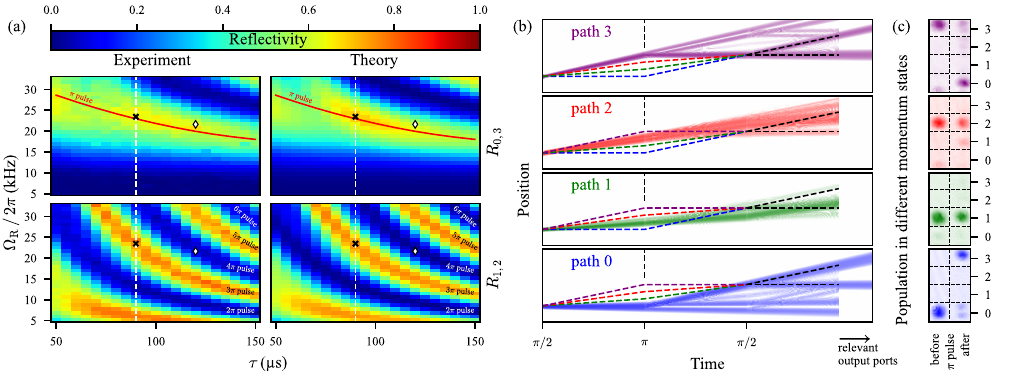}
    \caption{
    Experiment and simulation of third-order Bragg diffraction with DMPs.
    (a) Comparison of experimental (left) and numerical (right) reflectivities $R_{0,3}$ (top) for resonant six-photon and $R_{1,2}$ (bottom) for parasitic two-photon diffraction. 
    Scanning $\tau$ and $\Omega_\text{R}$ reveals different effective Rabi frequencies for both cases.
    The white diamonds indicate the DMP [$\tau=\SI{120}{\micro\second}$, $\Omega_\text{R}=2\pi\times\SI{21(2)}{\kilo\hertz}$] with significantly improved performance over the mirror pulse (black cross) obtained in \figref[a]{fig:init_sit}. 
    The labels indicate pulse areas of multiples of $\pi$.
    Panels (b) and (c) modify their counterparts of \figref{fig:init_sit} using the optimum DMP.
    In (b), resonant paths are reflected efficiently but parasitic paths are fully transmitted by the mirror pulse and thus not coupled into the relevant output ports, as highlighted by the lack of relevant population in the final section of the plot.
    (c) We confirm this dichroic behavior experimentally by absorption images of the momentum distribution before (left) and after (right) the DMP, where parasitic paths are not reflected by the DMP and almost fully remain in their input momentum class.
    }
    \label{fig:final_sit}
\end{figure*}

As in \figref[b]{fig:init_sit}, we display a path-resolved MZI simulation in \figref[b]{fig:final_sit}.
Since the DMP does not redirect  parasitic paths, they do not overlap with  resonant arms at the second beam splitter and are not coupled into the two relevant output ports. 
We verify this dichroic behavior experimentally. 
\figureref[c]{fig:final_sit} displays the momentum distributions before (left) and after (right) the DMP. 
Indeed, the DMP is reflective for resonant paths and redirects them to the exit ports, while maintaining near-perfect transparency for both parasitic paths where the output momentum distribution resembles the input.

While the concept of DMPs has been theoretically studied for momentum eigenstates~\cite{kirsten:2023:prl}, our experiments demonstrate its applicability to realistic momentum distributions. 
To  analyze its robustness, we numerically determine $R_{0,3}$ and $R_{1,2}$ as a function of the initial momentum spread $\Delta p$ in \figref[a]{fig:momentum_width}. 
For increasing $\Delta p$ we observe that $R_{0,3}$ deteriorates for both pulse settings and approaches unity for small $\Delta p$, as expected from velocity selectivity. 
However, the DMP exhibits markedly reduced reflectivity $R_{1,2}$ over the full range in $\Delta p$.
Hence, our simulations highlight that DMPs are operational for a broad range of momentum distributions. 
\begin{figure}[ht]
    \centering
    \includegraphics[width=1\linewidth]{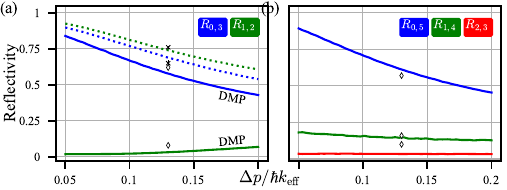}
    \caption{
    Simulated reflectivities (lines) as a function of momentum spread $\Delta p$ complemented by experimental values for $\Delta p = 0.13 \hbar k_\text{eff}$ (crosses and diamonds).
    (a) Third-order diffraction:
    Comparing a pulse with the parameters of maximal $R_{0,3}$ from \figref[a]{fig:init_sit} (dotted lines and black crosses) to the DMP (solid lines and white diamonds), we observe a drastic drop in the reflectivity of parasitic paths for the DMP for all $\Delta p$.
    (b) A DMP of fifth order shows analog behavior. 
    }
    \label{fig:momentum_width}
\end{figure}

Our results can be transferred to any odd diffraction order, in particular to fifth order, which is a good compromise~\cite{Szigeti:2012:momentumwidth} between available laser power, velocity selectivity, and loss from spontaneous emission. 
We experimentally demonstrate this scalabiltiy by implementing a fifth-order DMP and measuring $R_{0,5}$, $R_{1,4}$, and $R_{2,3}$ [\figref[b]{fig:fifth_order}].
The latter two reflectivities are associated with parasitic paths as shown in \figref[a]{fig:fifth_order}. 
The different scaling of the multi-photon Rabi frequencies~\cite{mueller:2008:atom-wave} allows to identify parameters [$\tau =\SI{100}{\micro\second}$, $\rabi=2\pi\times\SI{52(5)}{\kilo\hertz}$] where the pulse area is close to $\pi$ for the resonant path, but $4\pi$ and $6\pi$ for the parasitic ones, giving reflectivities  $R_{0,5}=0.57(1)$, $R_{1,4}=0.16(1)$, $R_{2,3}=0.10(1)$.  
In \figref[b]{fig:momentum_width} we simulate the influence of $\Delta p$ on the reflectivities, using $\rabi=2\pi\times\SI{52.9}{\kilo\hertz}$ which has been obtained from a fit to the Rabi scan of $R_{1,4}$ for $\tau =\SI{100}{\micro\second}$ and lies within the error margin of the experimental value reported above.
We again observe drastically reduced parasitic reflectivities for the DMP for all $\Delta p$.
\begin{figure}[ht]
    \centering
    \includegraphics[width=1\linewidth]{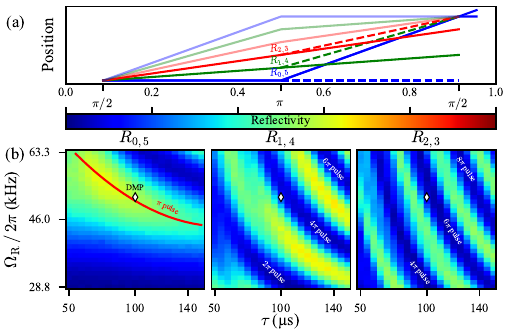}
    \caption{
    Fifth-order MZI (a) highlighting the paths associated with the reflectivities $R_{0,5}$, $R_{1,4}$, and $R_{2,3}$ as measured in (b) for different pulse lengths $\tau$ and two-photon Rabi frequencies $\rabi$.
    The DMP is marked by a white diamond ($\tau =\SI{100}{\micro\second}$, $\rabi=2\pi\times\SI{52(5)}{\kilo\hertz}$), where the resonant path experiences a pulse area close to $\pi$, while the parasitic paths undergo $4\pi$ and $6\pi$ pulses, respectively.
    }
    \label{fig:fifth_order}
\end{figure}

Additional simulations for $n=7$ and $n=9$  (with $\Delta p = 0.05 \hbar k_\text{eff}$) demonstrate that high reflectivity of the resonant and simultaneous suppression of the reflectivities for all parasitic orders is achievable, indicating that this technique scales to even higher diffraction orders. Note, that extending the scheme to even diffraction orders has the disadvantage that the central parasitic path will always couple into the output ports, similar to double diffraction~\cite{jenewein:2022:bragg}.
While we have focused on the reflectivity of resonant and parasitic paths in this Letter, it is possible to use different optimization strategies and more refined parameters that, e.\,g., include the population of parasitic orders after the initial beam splitter or to customize the method to geometries beyond MZIs~\cite{sidorenkov:2020:tailoring}. 
After all, the ultimate figure of merit is the sensitivity of a closed interferometer, which intrinsically depends on the contrast.
Using optimal control or reinforcement machine learning techniques might improve the performance of DMPs even further~\cite{chih:2021:reinforcement,Saywell:2020:optimalcontrol,saywell:2020:biselective}. 

Although we studied DMPs for higher-order Bragg diffraction, the concept may be transferred to other approaches of LMT like sequential pulses~\cite{rudolph:2020:largemomentum,berg:2015:composite-light,mcguirk:2000:largearea}, which can be also applied to other diffraction, e.\,g. clock transitions~\cite{rudolph:2020:largemomentum, hu:2017:atominterferometry}. 
Since composite pulses often resort to a combination of several LMT techniques, each with its own mitigation scheme for parasitic orders, the interplay of these strategies will eventually be of relevance for advanced experiments~\cite{abend:2016:atom,mueller:2008:atom,morel:2020:determination,parker:2018:measurement,ahlers:2016:doublebragg, mueller:2009:atom}. 
Adding DMPs to the toolbox of LMT will foster the implementation of ambitious proposals for high-precision atom interferometry. 

\begin{acknowledgments}
We thank L. Lind and the QUANTUS and INTENTAS teams for helpful discussions.
The QUANTUS project is supported by the German Space Agency at the German Aerospace Center (Deutsche Raumfahrtagentur im Deutschen Zentrum f\"ur Luft- und Raumfahrt, DLR) with funds provided by the Federal Ministry for Economic Affairs and Climate Action (Bundesministerium f\"ur Wirtschaft und Klimaschutz, BMWK) due to an enactment of the German Bundestag under Grant No. 50WM2450E (QUANTUS-VI) .
\end{acknowledgments}

\bibliography{Bibfile}

\begin{thebibliography}{74}%
\makeatletter
\providecommand \@ifxundefined [1]{%
 \@ifx{#1\undefined}
}%
\providecommand \@ifnum [1]{%
 \ifnum #1\expandafter \@firstoftwo
 \else \expandafter \@secondoftwo
 \fi
}%
\providecommand \@ifx [1]{%
 \ifx #1\expandafter \@firstoftwo
 \else \expandafter \@secondoftwo
 \fi
}%
\providecommand \natexlab [1]{#1}%
\providecommand \enquote  [1]{``#1''}%
\providecommand \bibnamefont  [1]{#1}%
\providecommand \bibfnamefont [1]{#1}%
\providecommand \citenamefont [1]{#1}%
\providecommand \href@noop [0]{\@secondoftwo}%
\providecommand \href [0]{\begingroup \@sanitize@url \@href}%
\providecommand \@href[1]{\@@startlink{#1}\@@href}%
\providecommand \@@href[1]{\endgroup#1\@@endlink}%
\providecommand \@sanitize@url [0]{\catcode `\\12\catcode `\$12\catcode
  `\&12\catcode `\#12\catcode `\^12\catcode `\_12\catcode `\%12\relax}%
\providecommand \@@startlink[1]{}%
\providecommand \@@endlink[0]{}%
\providecommand \url  [0]{\begingroup\@sanitize@url \@url }%
\providecommand \@url [1]{\endgroup\@href {#1}{\urlprefix }}%
\providecommand \urlprefix  [0]{URL }%
\providecommand \Eprint [0]{\href }%
\providecommand \doibase [0]{https://doi.org/}%
\providecommand \selectlanguage [0]{\@gobble}%
\providecommand \bibinfo  [0]{\@secondoftwo}%
\providecommand \bibfield  [0]{\@secondoftwo}%
\providecommand \translation [1]{[#1]}%
\providecommand \BibitemOpen [0]{}%
\providecommand \bibitemStop [0]{}%
\providecommand \bibitemNoStop [0]{.\EOS\space}%
\providecommand \EOS [0]{\spacefactor3000\relax}%
\providecommand \BibitemShut  [1]{\csname bibitem#1\endcsname}%
\let\auto@bib@innerbib\@empty
\bibitem [{\citenamefont {Debs}\ \emph {et~al.}(2011)\citenamefont {Debs},
  \citenamefont {Altin}, \citenamefont {Barter}, \citenamefont {D\"oring},
  \citenamefont {Dennis}, \citenamefont {McDonald}, \citenamefont {Anderson},
  \citenamefont {Close},\ and\ \citenamefont {Robins}}]{debs:2011:cold-atom}%
  \BibitemOpen
  \bibfield  {author} {\bibinfo {author} {\bibfnamefont {J.~E.}\ \bibnamefont
  {Debs}}, \bibinfo {author} {\bibfnamefont {P.~A.}\ \bibnamefont {Altin}},
  \bibinfo {author} {\bibfnamefont {T.~H.}\ \bibnamefont {Barter}}, \bibinfo
  {author} {\bibfnamefont {D.}~\bibnamefont {D\"oring}}, \bibinfo {author}
  {\bibfnamefont {G.~R.}\ \bibnamefont {Dennis}}, \bibinfo {author}
  {\bibfnamefont {G.}~\bibnamefont {McDonald}}, \bibinfo {author}
  {\bibfnamefont {R.~P.}\ \bibnamefont {Anderson}}, \bibinfo {author}
  {\bibfnamefont {J.~D.}\ \bibnamefont {Close}},\ and\ \bibinfo {author}
  {\bibfnamefont {N.~P.}\ \bibnamefont {Robins}},\ }\bibfield  {title}
  {\bibinfo {title} {{Cold-atom gravimetry with a Bose-Einstein condensate}},\
  }\href {https://doi.org/10.1103/PhysRevA.84.033610} {\bibfield  {journal}
  {\bibinfo  {journal} {Phys. Rev. A}\ }\textbf {\bibinfo {volume} {84}},\
  \bibinfo {pages} {033610} (\bibinfo {year} {2011})}\BibitemShut {NoStop}%
\bibitem [{\citenamefont {Chiow}\ \emph {et~al.}(2011)\citenamefont {Chiow},
  \citenamefont {Kovachy}, \citenamefont {Chien},\ and\ \citenamefont
  {Kasevich}}]{chiow:2011:102}%
  \BibitemOpen
  \bibfield  {author} {\bibinfo {author} {\bibfnamefont {S.-w.}\ \bibnamefont
  {Chiow}}, \bibinfo {author} {\bibfnamefont {T.}~\bibnamefont {Kovachy}},
  \bibinfo {author} {\bibfnamefont {H.-C.}\ \bibnamefont {Chien}},\ and\
  \bibinfo {author} {\bibfnamefont {M.~A.}\ \bibnamefont {Kasevich}},\
  }\bibfield  {title} {\bibinfo {title} {{$102\ensuremath{\hbar}k$ Large Area
  Atom Interferometers}},\ }\href
  {https://doi.org/10.1103/PhysRevLett.107.130403} {\bibfield  {journal}
  {\bibinfo  {journal} {Phys. Rev. Lett.}\ }\textbf {\bibinfo {volume} {107}},\
  \bibinfo {pages} {130403} (\bibinfo {year} {2011})}\BibitemShut {NoStop}%
\bibitem [{\citenamefont {M\"untinga}\ \emph {et~al.}(2013)\citenamefont
  {M\"untinga}, \citenamefont {Ahlers}, \citenamefont {Krutzik}, \citenamefont
  {Wenzlawski}, \citenamefont {Arnold}, \citenamefont {Becker}, \citenamefont
  {Bongs}, \citenamefont {Dittus}, \citenamefont {Duncker}, \citenamefont
  {Gaaloul}, \citenamefont {Gherasim}, \citenamefont {Giese}, \citenamefont
  {Grzeschik}, \citenamefont {H\"ansch}, \citenamefont {Hellmig}, \citenamefont
  {Herr}, \citenamefont {Herrmann}, \citenamefont {Kajari}, \citenamefont
  {Kleinert}, \citenamefont {L\"ammerzahl} \emph
  {et~al.}}]{muentinga:2013:interferometry}%
  \BibitemOpen
  \bibfield  {author} {\bibinfo {author} {\bibfnamefont {H.}~\bibnamefont
  {M\"untinga}}, \bibinfo {author} {\bibfnamefont {H.}~\bibnamefont {Ahlers}},
  \bibinfo {author} {\bibfnamefont {M.}~\bibnamefont {Krutzik}}, \bibinfo
  {author} {\bibfnamefont {A.}~\bibnamefont {Wenzlawski}}, \bibinfo {author}
  {\bibfnamefont {S.}~\bibnamefont {Arnold}}, \bibinfo {author} {\bibfnamefont
  {D.}~\bibnamefont {Becker}}, \bibinfo {author} {\bibfnamefont
  {K.}~\bibnamefont {Bongs}}, \bibinfo {author} {\bibfnamefont
  {H.}~\bibnamefont {Dittus}}, \bibinfo {author} {\bibfnamefont
  {H.}~\bibnamefont {Duncker}}, \bibinfo {author} {\bibfnamefont
  {N.}~\bibnamefont {Gaaloul}}, \bibinfo {author} {\bibfnamefont
  {C.}~\bibnamefont {Gherasim}}, \bibinfo {author} {\bibfnamefont
  {E.}~\bibnamefont {Giese}}, \bibinfo {author} {\bibfnamefont
  {C.}~\bibnamefont {Grzeschik}}, \bibinfo {author} {\bibfnamefont {T.~W.}\
  \bibnamefont {H\"ansch}}, \bibinfo {author} {\bibfnamefont {O.}~\bibnamefont
  {Hellmig}}, \bibinfo {author} {\bibfnamefont {W.}~\bibnamefont {Herr}},
  \bibinfo {author} {\bibfnamefont {S.}~\bibnamefont {Herrmann}}, \bibinfo
  {author} {\bibfnamefont {E.}~\bibnamefont {Kajari}}, \bibinfo {author}
  {\bibfnamefont {S.}~\bibnamefont {Kleinert}}, \bibinfo {author}
  {\bibfnamefont {C.}~\bibnamefont {L\"ammerzahl}}, \emph {et~al.},\ }\bibfield
   {title} {\bibinfo {title} {{Interferometry with Bose-Einstein Condensates in
  Microgravity}},\ }\href {https://doi.org/10.1103/PhysRevLett.110.093602}
  {\bibfield  {journal} {\bibinfo  {journal} {Phys. Rev. Lett.}\ }\textbf
  {\bibinfo {volume} {110}},\ \bibinfo {pages} {093602} (\bibinfo {year}
  {2013})}\BibitemShut {NoStop}%
\bibitem [{\citenamefont {Barrett}\ \emph {et~al.}(2016)\citenamefont
  {Barrett}, \citenamefont {Antoni-Micollier}, \citenamefont {Chichet},
  \citenamefont {Battelier}, \citenamefont {L{\'e}v{\`e}que}, \citenamefont
  {Landragin},\ and\ \citenamefont {Bouyer}}]{Barrett:2016:dual}%
  \BibitemOpen
  \bibfield  {author} {\bibinfo {author} {\bibfnamefont {B.}~\bibnamefont
  {Barrett}}, \bibinfo {author} {\bibfnamefont {L.}~\bibnamefont
  {Antoni-Micollier}}, \bibinfo {author} {\bibfnamefont {L.}~\bibnamefont
  {Chichet}}, \bibinfo {author} {\bibfnamefont {B.}~\bibnamefont {Battelier}},
  \bibinfo {author} {\bibfnamefont {T.}~\bibnamefont {L{\'e}v{\`e}que}},
  \bibinfo {author} {\bibfnamefont {A.}~\bibnamefont {Landragin}},\ and\
  \bibinfo {author} {\bibfnamefont {P.}~\bibnamefont {Bouyer}},\ }\bibfield
  {title} {\bibinfo {title} {{Dual matter-wave inertial sensors in
  weightlessness}},\ }\href {https://doi.org/10.1038/ncomms13786} {\bibfield
  {journal} {\bibinfo  {journal} {Nat. Commun.}\ }\textbf {\bibinfo {volume}
  {7}},\ \bibinfo {pages} {13786} (\bibinfo {year} {2016})}\BibitemShut
  {NoStop}%
\bibitem [{\citenamefont {Lachmann}\ \emph {et~al.}(2021)\citenamefont
  {Lachmann}, \citenamefont {Ahlers}, \citenamefont {Becker}, \citenamefont
  {Dinkelaker}, \citenamefont {Grosse}, \citenamefont {Hellmig}, \citenamefont
  {M{\"u}ntinga}, \citenamefont {Schkolnik}, \citenamefont {Seidel},
  \citenamefont {Wendrich}, \citenamefont {Wenzlawski}, \citenamefont
  {Carrick}, \citenamefont {Gaaloul}, \citenamefont {L{\"u}dtke}, \citenamefont
  {Braxmaier}, \citenamefont {Ertmer}, \citenamefont {Krutzik}, \citenamefont
  {L{\"a}mmerzahl}, \citenamefont {Peters}, \citenamefont {Schleich},
  \citenamefont {Sengstock}, \citenamefont {Wicht}, \citenamefont
  {Windpassinger},\ and\ \citenamefont {Rasel}}]{Lachmann:2021:ultracold}%
  \BibitemOpen
  \bibfield  {author} {\bibinfo {author} {\bibfnamefont {M.~D.}\ \bibnamefont
  {Lachmann}}, \bibinfo {author} {\bibfnamefont {H.}~\bibnamefont {Ahlers}},
  \bibinfo {author} {\bibfnamefont {D.}~\bibnamefont {Becker}}, \bibinfo
  {author} {\bibfnamefont {A.~N.}\ \bibnamefont {Dinkelaker}}, \bibinfo
  {author} {\bibfnamefont {J.}~\bibnamefont {Grosse}}, \bibinfo {author}
  {\bibfnamefont {O.}~\bibnamefont {Hellmig}}, \bibinfo {author} {\bibfnamefont
  {H.}~\bibnamefont {M{\"u}ntinga}}, \bibinfo {author} {\bibfnamefont
  {V.}~\bibnamefont {Schkolnik}}, \bibinfo {author} {\bibfnamefont {S.~T.}\
  \bibnamefont {Seidel}}, \bibinfo {author} {\bibfnamefont {T.}~\bibnamefont
  {Wendrich}}, \bibinfo {author} {\bibfnamefont {A.}~\bibnamefont
  {Wenzlawski}}, \bibinfo {author} {\bibfnamefont {B.}~\bibnamefont {Carrick}},
  \bibinfo {author} {\bibfnamefont {N.}~\bibnamefont {Gaaloul}}, \bibinfo
  {author} {\bibfnamefont {D.}~\bibnamefont {L{\"u}dtke}}, \bibinfo {author}
  {\bibfnamefont {C.}~\bibnamefont {Braxmaier}}, \bibinfo {author}
  {\bibfnamefont {W.}~\bibnamefont {Ertmer}}, \bibinfo {author} {\bibfnamefont
  {M.}~\bibnamefont {Krutzik}}, \bibinfo {author} {\bibfnamefont
  {C.}~\bibnamefont {L{\"a}mmerzahl}}, \bibinfo {author} {\bibfnamefont
  {A.}~\bibnamefont {Peters}}, \bibinfo {author} {\bibfnamefont {W.~P.}\
  \bibnamefont {Schleich}}, \bibinfo {author} {\bibfnamefont {K.}~\bibnamefont
  {Sengstock}}, \bibinfo {author} {\bibfnamefont {A.}~\bibnamefont {Wicht}},
  \bibinfo {author} {\bibfnamefont {P.}~\bibnamefont {Windpassinger}},\ and\
  \bibinfo {author} {\bibfnamefont {E.~M.}\ \bibnamefont {Rasel}},\ }\bibfield
  {title} {\bibinfo {title} {{Ultracold atom interferometry in space}},\ }\href
  {https://doi.org/10.1038/s41467-021-21628-z} {\bibfield  {journal} {\bibinfo
  {journal} {Nat. Commun.}\ }\textbf {\bibinfo {volume} {12}},\ \bibinfo
  {pages} {1317} (\bibinfo {year} {2021})}\BibitemShut {NoStop}%
\bibitem [{\citenamefont {Williams}\ \emph {et~al.}(2024)\citenamefont
  {Williams}, \citenamefont {Sackett}, \citenamefont {Ahlers}, \citenamefont
  {Aveline}, \citenamefont {Boegel}, \citenamefont {Botsi}, \citenamefont
  {Charron}, \citenamefont {Elliott}, \citenamefont {Gaaloul}, \citenamefont
  {Giese}, \citenamefont {Herr}, \citenamefont {Kellogg}, \citenamefont
  {Kohel}, \citenamefont {Lay}, \citenamefont {Meister}, \citenamefont
  {M{\"u}ller}, \citenamefont {M{\"u}ller}, \citenamefont {Oudrhiri},
  \citenamefont {Phillips}, \citenamefont {Pichery}, \citenamefont {Rasel},
  \citenamefont {Roura}, \citenamefont {Sbroscia}, \citenamefont {Schleich},
  \citenamefont {Schneider}, \citenamefont {Schubert}, \citenamefont {Sen},
  \citenamefont {Thompson},\ and\ \citenamefont
  {Bigelow}}]{williams:2024:pathfinder}%
  \BibitemOpen
  \bibfield  {author} {\bibinfo {author} {\bibfnamefont {J.~R.}\ \bibnamefont
  {Williams}}, \bibinfo {author} {\bibfnamefont {C.~A.}\ \bibnamefont
  {Sackett}}, \bibinfo {author} {\bibfnamefont {H.}~\bibnamefont {Ahlers}},
  \bibinfo {author} {\bibfnamefont {D.~C.}\ \bibnamefont {Aveline}}, \bibinfo
  {author} {\bibfnamefont {P.}~\bibnamefont {Boegel}}, \bibinfo {author}
  {\bibfnamefont {S.}~\bibnamefont {Botsi}}, \bibinfo {author} {\bibfnamefont
  {E.}~\bibnamefont {Charron}}, \bibinfo {author} {\bibfnamefont {E.~R.}\
  \bibnamefont {Elliott}}, \bibinfo {author} {\bibfnamefont {N.}~\bibnamefont
  {Gaaloul}}, \bibinfo {author} {\bibfnamefont {E.}~\bibnamefont {Giese}},
  \bibinfo {author} {\bibfnamefont {W.}~\bibnamefont {Herr}}, \bibinfo {author}
  {\bibfnamefont {J.~R.}\ \bibnamefont {Kellogg}}, \bibinfo {author}
  {\bibfnamefont {J.~M.}\ \bibnamefont {Kohel}}, \bibinfo {author}
  {\bibfnamefont {N.~E.}\ \bibnamefont {Lay}}, \bibinfo {author} {\bibfnamefont
  {M.}~\bibnamefont {Meister}}, \bibinfo {author} {\bibfnamefont
  {G.}~\bibnamefont {M{\"u}ller}}, \bibinfo {author} {\bibfnamefont
  {H.}~\bibnamefont {M{\"u}ller}}, \bibinfo {author} {\bibfnamefont
  {K.}~\bibnamefont {Oudrhiri}}, \bibinfo {author} {\bibfnamefont
  {L.}~\bibnamefont {Phillips}}, \bibinfo {author} {\bibfnamefont
  {A.}~\bibnamefont {Pichery}}, \bibinfo {author} {\bibfnamefont {E.~M.}\
  \bibnamefont {Rasel}}, \bibinfo {author} {\bibfnamefont {A.}~\bibnamefont
  {Roura}}, \bibinfo {author} {\bibfnamefont {M.}~\bibnamefont {Sbroscia}},
  \bibinfo {author} {\bibfnamefont {W.~P.}\ \bibnamefont {Schleich}}, \bibinfo
  {author} {\bibfnamefont {C.}~\bibnamefont {Schneider}}, \bibinfo {author}
  {\bibfnamefont {C.}~\bibnamefont {Schubert}}, \bibinfo {author}
  {\bibfnamefont {B.}~\bibnamefont {Sen}}, \bibinfo {author} {\bibfnamefont
  {R.~J.}\ \bibnamefont {Thompson}},\ and\ \bibinfo {author} {\bibfnamefont
  {N.~P.}\ \bibnamefont {Bigelow}},\ }\bibfield  {title} {\bibinfo {title}
  {{Pathfinder experiments with atom interferometry in the Cold Atom Lab
  onboard the International Space Station}},\ }\href
  {https://doi.org/10.1038/s41467-024-50585-6} {\bibfield  {journal} {\bibinfo
  {journal} {Nat. Commun.}\ }\textbf {\bibinfo {volume} {15}},\ \bibinfo
  {pages} {6414} (\bibinfo {year} {2024})}\BibitemShut {NoStop}%
\bibitem [{\citenamefont {Schlippert}\ \emph {et~al.}(2020)\citenamefont
  {Schlippert}, \citenamefont {Meiners}, \citenamefont {Rengelink},
  \citenamefont {Schubert}, \citenamefont {Tell}, \citenamefont {Wodey},
  \citenamefont {Zipfel}, \citenamefont {Ertmer},\ and\ \citenamefont
  {Rasel}}]{schlippert:2020:matterwave}%
  \BibitemOpen
  \bibfield  {author} {\bibinfo {author} {\bibfnamefont {D.}~\bibnamefont
  {Schlippert}}, \bibinfo {author} {\bibfnamefont {C.}~\bibnamefont {Meiners}},
  \bibinfo {author} {\bibfnamefont {R.}~\bibnamefont {Rengelink}}, \bibinfo
  {author} {\bibfnamefont {C.}~\bibnamefont {Schubert}}, \bibinfo {author}
  {\bibfnamefont {D.}~\bibnamefont {Tell}}, \bibinfo {author} {\bibfnamefont
  {{\'E}.}~\bibnamefont {Wodey}}, \bibinfo {author} {\bibfnamefont
  {K.}~\bibnamefont {Zipfel}}, \bibinfo {author} {\bibfnamefont
  {W.}~\bibnamefont {Ertmer}},\ and\ \bibinfo {author} {\bibfnamefont
  {E.}~\bibnamefont {Rasel}},\ }\bibinfo {title} {{Matter-Wave Interferometry
  for Inertial Sensing and Tests of Fundamental Physics}},\ in\ \href
  {https://doi.org/10.1142/9789811213984_0010} {\emph {\bibinfo {booktitle}
  {CPT and Lorentz Symmetry}}}\ (\bibinfo  {publisher} {World Scientific},\
  \bibinfo {year} {2020})\ pp.\ \bibinfo {pages} {37--40}\BibitemShut {NoStop}%
\bibitem [{\citenamefont {Badurina}\ \emph {et~al.}(2020)\citenamefont
  {Badurina}, \citenamefont {Bentine}, \citenamefont {Blas}, \citenamefont
  {Bongs}, \citenamefont {Bortoletto}, \citenamefont {Bowcock}, \citenamefont
  {Bridges}, \citenamefont {Bowden}, \citenamefont {Buchmueller}, \citenamefont
  {Burrage}, \citenamefont {Coleman}, \citenamefont {Elertas}, \citenamefont
  {Ellis}, \citenamefont {Foot}, \citenamefont {Gibson}, \citenamefont
  {Haehnelt}, \citenamefont {Harte}, \citenamefont {Hedges}, \citenamefont
  {Hobson}, \citenamefont {Holynski} \emph {et~al.}}]{Badurina:2020:aion}%
  \BibitemOpen
  \bibfield  {author} {\bibinfo {author} {\bibfnamefont {L.}~\bibnamefont
  {Badurina}}, \bibinfo {author} {\bibfnamefont {E.}~\bibnamefont {Bentine}},
  \bibinfo {author} {\bibfnamefont {D.}~\bibnamefont {Blas}}, \bibinfo {author}
  {\bibfnamefont {K.}~\bibnamefont {Bongs}}, \bibinfo {author} {\bibfnamefont
  {D.}~\bibnamefont {Bortoletto}}, \bibinfo {author} {\bibfnamefont
  {T.}~\bibnamefont {Bowcock}}, \bibinfo {author} {\bibfnamefont
  {K.}~\bibnamefont {Bridges}}, \bibinfo {author} {\bibfnamefont
  {W.}~\bibnamefont {Bowden}}, \bibinfo {author} {\bibfnamefont
  {O.}~\bibnamefont {Buchmueller}}, \bibinfo {author} {\bibfnamefont
  {C.}~\bibnamefont {Burrage}}, \bibinfo {author} {\bibfnamefont
  {J.}~\bibnamefont {Coleman}}, \bibinfo {author} {\bibfnamefont
  {G.}~\bibnamefont {Elertas}}, \bibinfo {author} {\bibfnamefont
  {J.}~\bibnamefont {Ellis}}, \bibinfo {author} {\bibfnamefont
  {C.}~\bibnamefont {Foot}}, \bibinfo {author} {\bibfnamefont {V.}~\bibnamefont
  {Gibson}}, \bibinfo {author} {\bibfnamefont {M.}~\bibnamefont {Haehnelt}},
  \bibinfo {author} {\bibfnamefont {T.}~\bibnamefont {Harte}}, \bibinfo
  {author} {\bibfnamefont {S.}~\bibnamefont {Hedges}}, \bibinfo {author}
  {\bibfnamefont {R.}~\bibnamefont {Hobson}}, \bibinfo {author} {\bibfnamefont
  {M.}~\bibnamefont {Holynski}}, \emph {et~al.},\ }\bibfield  {title} {\bibinfo
  {title} {{AION: an atom interferometer observatory and network}},\ }\href
  {https://doi.org/10.1088/1475-7516/2020/05/011} {\bibfield  {journal}
  {\bibinfo  {journal} {J. Cosmol. Astropart. Phys.}\ }\textbf {\bibinfo
  {volume} {2020}}\bibinfo  {number} { (05)},\ \bibinfo {pages}
  {011}}\BibitemShut {NoStop}%
\bibitem [{\citenamefont {Zhou}\ \emph {et~al.}(2011)\citenamefont {Zhou},
  \citenamefont {Xiong}, \citenamefont {Yang}, \citenamefont {Tang},
  \citenamefont {Peng}, \citenamefont {Hao}, \citenamefont {Li}, \citenamefont
  {Liu}, \citenamefont {Wang},\ and\ \citenamefont
  {Zhan}}]{Zhou:2011:development}%
  \BibitemOpen
\bibfield  {number} {  }\bibfield  {author} {\bibinfo {author} {\bibfnamefont
  {L.}~\bibnamefont {Zhou}}, \bibinfo {author} {\bibfnamefont {Z.~Y.}\
  \bibnamefont {Xiong}}, \bibinfo {author} {\bibfnamefont {W.}~\bibnamefont
  {Yang}}, \bibinfo {author} {\bibfnamefont {B.}~\bibnamefont {Tang}}, \bibinfo
  {author} {\bibfnamefont {W.~C.}\ \bibnamefont {Peng}}, \bibinfo {author}
  {\bibfnamefont {K.}~\bibnamefont {Hao}}, \bibinfo {author} {\bibfnamefont
  {R.~B.}\ \bibnamefont {Li}}, \bibinfo {author} {\bibfnamefont
  {M.}~\bibnamefont {Liu}}, \bibinfo {author} {\bibfnamefont {J.}~\bibnamefont
  {Wang}},\ and\ \bibinfo {author} {\bibfnamefont {M.~S.}\ \bibnamefont
  {Zhan}},\ }\bibfield  {title} {\bibinfo {title} {{Development of an atom
  gravimeter and status of the 10-meter atom interferometer for precision
  gravity measurement}},\ }\href {https://doi.org/10.1007/s10714-011-1167-9}
  {\bibfield  {journal} {\bibinfo  {journal} {Gen. Relativ. Gravitation}\
  }\textbf {\bibinfo {volume} {43}},\ \bibinfo {pages} {1931} (\bibinfo {year}
  {2011})}\BibitemShut {NoStop}%
\bibitem [{\citenamefont {Kovachy}\ \emph {et~al.}(2015)\citenamefont
  {Kovachy}, \citenamefont {Asenbaum}, \citenamefont {Overstreet},
  \citenamefont {Donnelly}, \citenamefont {Dickerson}, \citenamefont
  {Sugarbaker}, \citenamefont {Hogan},\ and\ \citenamefont
  {Kasevich}}]{kovachy:2015:quantum}%
  \BibitemOpen
  \bibfield  {author} {\bibinfo {author} {\bibfnamefont {T.}~\bibnamefont
  {Kovachy}}, \bibinfo {author} {\bibfnamefont {P.}~\bibnamefont {Asenbaum}},
  \bibinfo {author} {\bibfnamefont {C.}~\bibnamefont {Overstreet}}, \bibinfo
  {author} {\bibfnamefont {C.~A.}\ \bibnamefont {Donnelly}}, \bibinfo {author}
  {\bibfnamefont {S.~M.}\ \bibnamefont {Dickerson}}, \bibinfo {author}
  {\bibfnamefont {A.}~\bibnamefont {Sugarbaker}}, \bibinfo {author}
  {\bibfnamefont {J.~M.}\ \bibnamefont {Hogan}},\ and\ \bibinfo {author}
  {\bibfnamefont {M.~A.}\ \bibnamefont {Kasevich}},\ }\bibfield  {title}
  {\bibinfo {title} {{Quantum superposition at the half-metre scale}},\ }\href
  {https://doi.org/10.1038/nature16155} {\bibfield  {journal} {\bibinfo
  {journal} {Nature}\ }\textbf {\bibinfo {volume} {528}},\ \bibinfo {pages}
  {530} (\bibinfo {year} {2015})}\BibitemShut {NoStop}%
\bibitem [{\citenamefont {Impens}\ and\ \citenamefont
  {Bord\'e}(2009)}]{impens:2009:spacetime}%
  \BibitemOpen
  \bibfield  {author} {\bibinfo {author} {\bibfnamefont {F.}~\bibnamefont
  {Impens}}\ and\ \bibinfo {author} {\bibfnamefont {C.~J.}\ \bibnamefont
  {Bord\'e}},\ }\bibfield  {title} {\bibinfo {title} {{Space-time sensors using
  multiple-wave atom levitation}},\ }\href
  {https://doi.org/10.1103/PhysRevA.80.031602} {\bibfield  {journal} {\bibinfo
  {journal} {Phys. Rev. A}\ }\textbf {\bibinfo {volume} {80}},\ \bibinfo
  {pages} {031602} (\bibinfo {year} {2009})}\BibitemShut {NoStop}%
\bibitem [{\citenamefont {Schubert}\ \emph {et~al.}(2021)\citenamefont
  {Schubert}, \citenamefont {Abend}, \citenamefont {Gersemann}, \citenamefont
  {Gebbe}, \citenamefont {Schlippert}, \citenamefont {Berg},\ and\
  \citenamefont {Rasel}}]{Schubert:2021:multi-loop}%
  \BibitemOpen
  \bibfield  {author} {\bibinfo {author} {\bibfnamefont {C.}~\bibnamefont
  {Schubert}}, \bibinfo {author} {\bibfnamefont {S.}~\bibnamefont {Abend}},
  \bibinfo {author} {\bibfnamefont {M.}~\bibnamefont {Gersemann}}, \bibinfo
  {author} {\bibfnamefont {M.}~\bibnamefont {Gebbe}}, \bibinfo {author}
  {\bibfnamefont {D.}~\bibnamefont {Schlippert}}, \bibinfo {author}
  {\bibfnamefont {P.}~\bibnamefont {Berg}},\ and\ \bibinfo {author}
  {\bibfnamefont {E.~M.}\ \bibnamefont {Rasel}},\ }\bibfield  {title} {\bibinfo
  {title} {{Multi-loop atomic Sagnac interferometry}},\ }\href
  {https://doi.org/10.1038/s41598-021-95334-7} {\bibfield  {journal} {\bibinfo
  {journal} {Sci. Rep.}\ }\textbf {\bibinfo {volume} {11}},\ \bibinfo {pages}
  {16121} (\bibinfo {year} {2021})}\BibitemShut {NoStop}%
\bibitem [{\citenamefont {Schubert}\ \emph {et~al.}(2024)\citenamefont
  {Schubert}, \citenamefont {Schlippert}, \citenamefont {Gersemann},
  \citenamefont {Abend}, \citenamefont {Giese} \emph
  {et~al.}}]{schubert:2019:scalable}%
  \BibitemOpen
  \bibfield  {author} {\bibinfo {author} {\bibfnamefont {C.}~\bibnamefont
  {Schubert}}, \bibinfo {author} {\bibfnamefont {D.}~\bibnamefont
  {Schlippert}}, \bibinfo {author} {\bibfnamefont {M.}~\bibnamefont
  {Gersemann}}, \bibinfo {author} {\bibfnamefont {S.}~\bibnamefont {Abend}},
  \bibinfo {author} {\bibfnamefont {E.}~\bibnamefont {Giese}}, \emph {et~al.},\
  }\bibfield  {title} {\bibinfo {title} {A scalable, symmetric atom
  interferometer for infrasound gravitational wave detection},\ }\href
  {https://doi.org/10.1116/5.0228398} {\bibfield  {journal} {\bibinfo
  {journal} {AVS Quantum Sci.}\ }\textbf {\bibinfo {volume} {6}},\ \bibinfo
  {pages} {044404} (\bibinfo {year} {2024})}\BibitemShut {NoStop}%
\bibitem [{\citenamefont {Abend}\ \emph {et~al.}(2016)\citenamefont {Abend},
  \citenamefont {Gebbe}, \citenamefont {Gersemann}, \citenamefont {Ahlers},
  \citenamefont {M\"untinga}, \citenamefont {Giese}, \citenamefont {Gaaloul},
  \citenamefont {Schubert}, \citenamefont {L\"ammerzahl}, \citenamefont
  {Ertmer}, \citenamefont {Schleich},\ and\ \citenamefont
  {Rasel}}]{abend:2016:atom}%
  \BibitemOpen
  \bibfield  {author} {\bibinfo {author} {\bibfnamefont {S.}~\bibnamefont
  {Abend}}, \bibinfo {author} {\bibfnamefont {M.}~\bibnamefont {Gebbe}},
  \bibinfo {author} {\bibfnamefont {M.}~\bibnamefont {Gersemann}}, \bibinfo
  {author} {\bibfnamefont {H.}~\bibnamefont {Ahlers}}, \bibinfo {author}
  {\bibfnamefont {H.}~\bibnamefont {M\"untinga}}, \bibinfo {author}
  {\bibfnamefont {E.}~\bibnamefont {Giese}}, \bibinfo {author} {\bibfnamefont
  {N.}~\bibnamefont {Gaaloul}}, \bibinfo {author} {\bibfnamefont
  {C.}~\bibnamefont {Schubert}}, \bibinfo {author} {\bibfnamefont
  {C.}~\bibnamefont {L\"ammerzahl}}, \bibinfo {author} {\bibfnamefont
  {W.}~\bibnamefont {Ertmer}}, \bibinfo {author} {\bibfnamefont {W.~P.}\
  \bibnamefont {Schleich}},\ and\ \bibinfo {author} {\bibfnamefont {E.~M.}\
  \bibnamefont {Rasel}},\ }\bibfield  {title} {\bibinfo {title} {{Atom-Chip
  Fountain Gravimeter}},\ }\href
  {https://doi.org/10.1103/PhysRevLett.117.203003} {\bibfield  {journal}
  {\bibinfo  {journal} {Phys. Rev. Lett.}\ }\textbf {\bibinfo {volume} {117}},\
  \bibinfo {pages} {203003} (\bibinfo {year} {2016})}\BibitemShut {NoStop}%
\bibitem [{\citenamefont {Hughes}\ \emph {et~al.}(2009)\citenamefont {Hughes},
  \citenamefont {Burke},\ and\ \citenamefont
  {Sackett}}]{hughes:2009:suspension}%
  \BibitemOpen
  \bibfield  {author} {\bibinfo {author} {\bibfnamefont {K.~J.}\ \bibnamefont
  {Hughes}}, \bibinfo {author} {\bibfnamefont {J.~H.~T.}\ \bibnamefont
  {Burke}},\ and\ \bibinfo {author} {\bibfnamefont {C.~A.}\ \bibnamefont
  {Sackett}},\ }\bibfield  {title} {\bibinfo {title} {{Suspension of Atoms
  Using Optical Pulses, and Application to Gravimetry}},\ }\href
  {https://doi.org/10.1103/PhysRevLett.102.150403} {\bibfield  {journal}
  {\bibinfo  {journal} {Phys. Rev. Lett.}\ }\textbf {\bibinfo {volume} {102}},\
  \bibinfo {pages} {150403} (\bibinfo {year} {2009})}\BibitemShut {NoStop}%
\bibitem [{\citenamefont {Morel}\ \emph {et~al.}(2020)\citenamefont {Morel},
  \citenamefont {Yao}, \citenamefont {Clad{\'e}},\ and\ \citenamefont
  {Guellati-Kh{\'e}lifa}}]{morel:2020:determination}%
  \BibitemOpen
  \bibfield  {author} {\bibinfo {author} {\bibfnamefont {L.}~\bibnamefont
  {Morel}}, \bibinfo {author} {\bibfnamefont {Z.}~\bibnamefont {Yao}}, \bibinfo
  {author} {\bibfnamefont {P.}~\bibnamefont {Clad{\'e}}},\ and\ \bibinfo
  {author} {\bibfnamefont {S.}~\bibnamefont {Guellati-Kh{\'e}lifa}},\
  }\bibfield  {title} {\bibinfo {title} {{Determination of the fine-structure
  constant with an accuracy of 81 parts per trillion}},\ }\href
  {https://doi.org/10.1038/s41586-020-2964-7} {\bibfield  {journal} {\bibinfo
  {journal} {Nature}\ }\textbf {\bibinfo {volume} {588}},\ \bibinfo {pages}
  {61} (\bibinfo {year} {2020})}\BibitemShut {NoStop}%
\bibitem [{\citenamefont {Parker}\ \emph {et~al.}(2018)\citenamefont {Parker},
  \citenamefont {Yu}, \citenamefont {Zhong}, \citenamefont {Estey},\ and\
  \citenamefont {Müller}}]{parker:2018:measurement}%
  \BibitemOpen
  \bibfield  {author} {\bibinfo {author} {\bibfnamefont {R.~H.}\ \bibnamefont
  {Parker}}, \bibinfo {author} {\bibfnamefont {C.}~\bibnamefont {Yu}}, \bibinfo
  {author} {\bibfnamefont {W.}~\bibnamefont {Zhong}}, \bibinfo {author}
  {\bibfnamefont {B.}~\bibnamefont {Estey}},\ and\ \bibinfo {author}
  {\bibfnamefont {H.}~\bibnamefont {Müller}},\ }\bibfield  {title} {\bibinfo
  {title} {{Measurement of the fine-structure constant as a test of the
  Standard Model}},\ }\href {https://doi.org/10.1126/science.aap7706}
  {\bibfield  {journal} {\bibinfo  {journal} {Science}\ }\textbf {\bibinfo
  {volume} {360}},\ \bibinfo {pages} {191} (\bibinfo {year}
  {2018})}\BibitemShut {NoStop}%
\bibitem [{\citenamefont {Asenbaum}\ \emph {et~al.}(2020)\citenamefont
  {Asenbaum}, \citenamefont {Overstreet}, \citenamefont {Kim}, \citenamefont
  {Curti},\ and\ \citenamefont
  {Kasevich}}]{asenbaum:2020:atom-interferometric}%
  \BibitemOpen
  \bibfield  {author} {\bibinfo {author} {\bibfnamefont {P.}~\bibnamefont
  {Asenbaum}}, \bibinfo {author} {\bibfnamefont {C.}~\bibnamefont
  {Overstreet}}, \bibinfo {author} {\bibfnamefont {M.}~\bibnamefont {Kim}},
  \bibinfo {author} {\bibfnamefont {J.}~\bibnamefont {Curti}},\ and\ \bibinfo
  {author} {\bibfnamefont {M.~A.}\ \bibnamefont {Kasevich}},\ }\bibfield
  {title} {\bibinfo {title} {{Atom-Interferometric Test of the Equivalence
  Principle at the $10^{\ensuremath{-}12}$ Level}},\ }\href
  {https://doi.org/10.1103/PhysRevLett.125.191101} {\bibfield  {journal}
  {\bibinfo  {journal} {Phys. Rev. Lett.}\ }\textbf {\bibinfo {volume} {125}},\
  \bibinfo {pages} {191101} (\bibinfo {year} {2020})}\BibitemShut {NoStop}%
\bibitem [{\citenamefont {Gebbe}\ \emph {et~al.}(2021)\citenamefont {Gebbe},
  \citenamefont {Siem{\ss}}, \citenamefont {Gersemann}, \citenamefont
  {M{\"u}ntinga}, \citenamefont {Herrmann}, \citenamefont {L{\"a}mmerzahl},
  \citenamefont {Ahlers}, \citenamefont {Gaaloul}, \citenamefont {Schubert},
  \citenamefont {Hammerer}, \citenamefont {Abend},\ and\ \citenamefont
  {Rasel}}]{gebbe:2021:twin}%
  \BibitemOpen
  \bibfield  {author} {\bibinfo {author} {\bibfnamefont {M.}~\bibnamefont
  {Gebbe}}, \bibinfo {author} {\bibfnamefont {J.-N.}\ \bibnamefont
  {Siem{\ss}}}, \bibinfo {author} {\bibfnamefont {M.}~\bibnamefont
  {Gersemann}}, \bibinfo {author} {\bibfnamefont {H.}~\bibnamefont
  {M{\"u}ntinga}}, \bibinfo {author} {\bibfnamefont {S.}~\bibnamefont
  {Herrmann}}, \bibinfo {author} {\bibfnamefont {C.}~\bibnamefont
  {L{\"a}mmerzahl}}, \bibinfo {author} {\bibfnamefont {H.}~\bibnamefont
  {Ahlers}}, \bibinfo {author} {\bibfnamefont {N.}~\bibnamefont {Gaaloul}},
  \bibinfo {author} {\bibfnamefont {C.}~\bibnamefont {Schubert}}, \bibinfo
  {author} {\bibfnamefont {K.}~\bibnamefont {Hammerer}}, \bibinfo {author}
  {\bibfnamefont {S.}~\bibnamefont {Abend}},\ and\ \bibinfo {author}
  {\bibfnamefont {E.~M.}\ \bibnamefont {Rasel}},\ }\bibfield  {title} {\bibinfo
  {title} {{Twin-lattice atom interferometry}},\ }\href
  {https://doi.org/10.1038/s41467-021-22823-8} {\bibfield  {journal} {\bibinfo
  {journal} {Nat. Commun.}\ }\textbf {\bibinfo {volume} {12}},\ \bibinfo
  {pages} {2544} (\bibinfo {year} {2021})}\BibitemShut {NoStop}%
\bibitem [{\citenamefont {Canuel}\ \emph {et~al.}(2020)\citenamefont {Canuel},
  \citenamefont {Abend}, \citenamefont {Amaro-Seoane}, \citenamefont
  {Badaracco}, \citenamefont {Beaufils}, \citenamefont {Bertoldi},
  \citenamefont {Bongs}, \citenamefont {Bouyer}, \citenamefont {Braxmaier},
  \citenamefont {Chaibi}, \citenamefont {Christensen}, \citenamefont {Fitzek},
  \citenamefont {Flouris}, \citenamefont {Gaaloul}, \citenamefont {Gaffet},
  \citenamefont {Garrido~Alzar}, \citenamefont {Geiger}, \citenamefont
  {Guellati-Khelifa}, \citenamefont {Hammerer}, \citenamefont {Harms} \emph
  {et~al.}}]{Canuel:2020:elgara}%
  \BibitemOpen
  \bibfield  {author} {\bibinfo {author} {\bibfnamefont {B.}~\bibnamefont
  {Canuel}}, \bibinfo {author} {\bibfnamefont {S.}~\bibnamefont {Abend}},
  \bibinfo {author} {\bibfnamefont {P.}~\bibnamefont {Amaro-Seoane}}, \bibinfo
  {author} {\bibfnamefont {F.}~\bibnamefont {Badaracco}}, \bibinfo {author}
  {\bibfnamefont {Q.}~\bibnamefont {Beaufils}}, \bibinfo {author}
  {\bibfnamefont {A.}~\bibnamefont {Bertoldi}}, \bibinfo {author}
  {\bibfnamefont {K.}~\bibnamefont {Bongs}}, \bibinfo {author} {\bibfnamefont
  {P.}~\bibnamefont {Bouyer}}, \bibinfo {author} {\bibfnamefont
  {C.}~\bibnamefont {Braxmaier}}, \bibinfo {author} {\bibfnamefont
  {W.}~\bibnamefont {Chaibi}}, \bibinfo {author} {\bibfnamefont
  {N.}~\bibnamefont {Christensen}}, \bibinfo {author} {\bibfnamefont
  {F.}~\bibnamefont {Fitzek}}, \bibinfo {author} {\bibfnamefont
  {G.}~\bibnamefont {Flouris}}, \bibinfo {author} {\bibfnamefont
  {N.}~\bibnamefont {Gaaloul}}, \bibinfo {author} {\bibfnamefont
  {S.}~\bibnamefont {Gaffet}}, \bibinfo {author} {\bibfnamefont {C.~L.}\
  \bibnamefont {Garrido~Alzar}}, \bibinfo {author} {\bibfnamefont
  {R.}~\bibnamefont {Geiger}}, \bibinfo {author} {\bibfnamefont
  {S.}~\bibnamefont {Guellati-Khelifa}}, \bibinfo {author} {\bibfnamefont
  {K.}~\bibnamefont {Hammerer}}, \bibinfo {author} {\bibfnamefont
  {J.}~\bibnamefont {Harms}}, \emph {et~al.},\ }\bibfield  {title} {\bibinfo
  {title} {{ELGAR—a European Laboratory for Gravitation and
  Atom-interferometric Research}},\ }\href
  {https://doi.org/10.1088/1361-6382/aba80e} {\bibfield  {journal} {\bibinfo
  {journal} {Classical Quantum Gravity}\ }\textbf {\bibinfo {volume} {37}},\
  \bibinfo {pages} {225017} (\bibinfo {year} {2020})}\BibitemShut {NoStop}%
\bibitem [{\citenamefont {Zhan}\ \emph {et~al.}(2020)\citenamefont {Zhan},
  \citenamefont {Wang}, \citenamefont {Ni}, \citenamefont {Gao}, \citenamefont
  {Wang}, \citenamefont {He}, \citenamefont {Li}, \citenamefont {Zhou},
  \citenamefont {Chen}, \citenamefont {Zhong}, \citenamefont {Tang},
  \citenamefont {Yao}, \citenamefont {Zhu}, \citenamefont {Xiong},
  \citenamefont {Lu}, \citenamefont {Yu}, \citenamefont {Cheng}, \citenamefont
  {Liu}, \citenamefont {Liang}, \citenamefont {Xu} \emph
  {et~al.}}]{zhan:2020:zaiga}%
  \BibitemOpen
  \bibfield  {author} {\bibinfo {author} {\bibfnamefont {M.-S.}\ \bibnamefont
  {Zhan}}, \bibinfo {author} {\bibfnamefont {J.}~\bibnamefont {Wang}}, \bibinfo
  {author} {\bibfnamefont {W.-T.}\ \bibnamefont {Ni}}, \bibinfo {author}
  {\bibfnamefont {D.-F.}\ \bibnamefont {Gao}}, \bibinfo {author} {\bibfnamefont
  {G.}~\bibnamefont {Wang}}, \bibinfo {author} {\bibfnamefont {L.-X.}\
  \bibnamefont {He}}, \bibinfo {author} {\bibfnamefont {R.-B.}\ \bibnamefont
  {Li}}, \bibinfo {author} {\bibfnamefont {L.}~\bibnamefont {Zhou}}, \bibinfo
  {author} {\bibfnamefont {X.}~\bibnamefont {Chen}}, \bibinfo {author}
  {\bibfnamefont {J.-Q.}\ \bibnamefont {Zhong}}, \bibinfo {author}
  {\bibfnamefont {B.}~\bibnamefont {Tang}}, \bibinfo {author} {\bibfnamefont
  {Z.-W.}\ \bibnamefont {Yao}}, \bibinfo {author} {\bibfnamefont
  {L.}~\bibnamefont {Zhu}}, \bibinfo {author} {\bibfnamefont {Z.-Y.}\
  \bibnamefont {Xiong}}, \bibinfo {author} {\bibfnamefont {S.-B.}\ \bibnamefont
  {Lu}}, \bibinfo {author} {\bibfnamefont {G.-H.}\ \bibnamefont {Yu}}, \bibinfo
  {author} {\bibfnamefont {Q.-F.}\ \bibnamefont {Cheng}}, \bibinfo {author}
  {\bibfnamefont {M.}~\bibnamefont {Liu}}, \bibinfo {author} {\bibfnamefont
  {Y.-R.}\ \bibnamefont {Liang}}, \bibinfo {author} {\bibfnamefont
  {P.}~\bibnamefont {Xu}}, \emph {et~al.},\ }\bibfield  {title} {\bibinfo
  {title} {{ZAIGA: Zhaoshan long-baseline atom interferometer gravitation
  antenna}},\ }\href {https://doi.org/10.1142/S0218271819400054} {\bibfield
  {journal} {\bibinfo  {journal} {Int. J. Mod. Phys. D}\ }\textbf {\bibinfo
  {volume} {29}},\ \bibinfo {pages} {1940005} (\bibinfo {year}
  {2020})}\BibitemShut {NoStop}%
\bibitem [{\citenamefont {Abend}\ \emph {et~al.}(2024)\citenamefont {Abend},
  \citenamefont {Allard}, \citenamefont {Alonso}, \citenamefont {Antoniadis},
  \citenamefont {Araújo}, \citenamefont {Arduini}, \citenamefont {Arnold},
  \citenamefont {Asano}, \citenamefont {Augst}, \citenamefont {Badurina},
  \citenamefont {Balaž}, \citenamefont {Banks}, \citenamefont {Barone},
  \citenamefont {Barsanti}, \citenamefont {Bassi}, \citenamefont {Battelier},
  \citenamefont {Baynham}, \citenamefont {Beaufils}, \citenamefont {Belić},
  \citenamefont {Beniwal} \emph {et~al.}}]{abend:2024:terrestrial}%
  \BibitemOpen
  \bibfield  {author} {\bibinfo {author} {\bibfnamefont {S.}~\bibnamefont
  {Abend}}, \bibinfo {author} {\bibfnamefont {B.}~\bibnamefont {Allard}},
  \bibinfo {author} {\bibfnamefont {I.}~\bibnamefont {Alonso}}, \bibinfo
  {author} {\bibfnamefont {J.}~\bibnamefont {Antoniadis}}, \bibinfo {author}
  {\bibfnamefont {H.}~\bibnamefont {Araújo}}, \bibinfo {author} {\bibfnamefont
  {G.}~\bibnamefont {Arduini}}, \bibinfo {author} {\bibfnamefont {A.~S.}\
  \bibnamefont {Arnold}}, \bibinfo {author} {\bibfnamefont {T.}~\bibnamefont
  {Asano}}, \bibinfo {author} {\bibfnamefont {N.}~\bibnamefont {Augst}},
  \bibinfo {author} {\bibfnamefont {L.}~\bibnamefont {Badurina}}, \bibinfo
  {author} {\bibfnamefont {A.}~\bibnamefont {Balaž}}, \bibinfo {author}
  {\bibfnamefont {H.}~\bibnamefont {Banks}}, \bibinfo {author} {\bibfnamefont
  {M.}~\bibnamefont {Barone}}, \bibinfo {author} {\bibfnamefont
  {M.}~\bibnamefont {Barsanti}}, \bibinfo {author} {\bibfnamefont
  {A.}~\bibnamefont {Bassi}}, \bibinfo {author} {\bibfnamefont
  {B.}~\bibnamefont {Battelier}}, \bibinfo {author} {\bibfnamefont {C.~F.~A.}\
  \bibnamefont {Baynham}}, \bibinfo {author} {\bibfnamefont {Q.}~\bibnamefont
  {Beaufils}}, \bibinfo {author} {\bibfnamefont {A.}~\bibnamefont {Belić}},
  \bibinfo {author} {\bibfnamefont {A.}~\bibnamefont {Beniwal}}, \emph
  {et~al.},\ }\bibfield  {title} {\bibinfo {title} {{Terrestrial
  very-long-baseline atom interferometry: Workshop summary}},\ }\href
  {https://doi.org/10.1116/5.0185291} {\bibfield  {journal} {\bibinfo
  {journal} {AVS Quantum Sci.}\ }\textbf {\bibinfo {volume} {6}},\ \bibinfo
  {pages} {024701} (\bibinfo {year} {2024})}\BibitemShut {NoStop}%
\bibitem [{\citenamefont {Abe}\ \emph {et~al.}(2021)\citenamefont {Abe},
  \citenamefont {Adamson}, \citenamefont {Borcean}, \citenamefont {Bortoletto},
  \citenamefont {Bridges}, \citenamefont {Carman}, \citenamefont
  {Chattopadhyay}, \citenamefont {Coleman}, \citenamefont {Curfman},
  \citenamefont {DeRose}, \citenamefont {Deshpande}, \citenamefont
  {Dimopoulos}, \citenamefont {Foot}, \citenamefont {Frisch}, \citenamefont
  {Garber}, \citenamefont {Geer}, \citenamefont {Gibson}, \citenamefont
  {Glick}, \citenamefont {Graham}, \citenamefont {Hahn} \emph
  {et~al.}}]{Abe:2021:matterwave}%
  \BibitemOpen
  \bibfield  {author} {\bibinfo {author} {\bibfnamefont {M.}~\bibnamefont
  {Abe}}, \bibinfo {author} {\bibfnamefont {P.}~\bibnamefont {Adamson}},
  \bibinfo {author} {\bibfnamefont {M.}~\bibnamefont {Borcean}}, \bibinfo
  {author} {\bibfnamefont {D.}~\bibnamefont {Bortoletto}}, \bibinfo {author}
  {\bibfnamefont {K.}~\bibnamefont {Bridges}}, \bibinfo {author} {\bibfnamefont
  {S.~P.}\ \bibnamefont {Carman}}, \bibinfo {author} {\bibfnamefont
  {S.}~\bibnamefont {Chattopadhyay}}, \bibinfo {author} {\bibfnamefont
  {J.}~\bibnamefont {Coleman}}, \bibinfo {author} {\bibfnamefont {N.~M.}\
  \bibnamefont {Curfman}}, \bibinfo {author} {\bibfnamefont {K.}~\bibnamefont
  {DeRose}}, \bibinfo {author} {\bibfnamefont {T.}~\bibnamefont {Deshpande}},
  \bibinfo {author} {\bibfnamefont {S.}~\bibnamefont {Dimopoulos}}, \bibinfo
  {author} {\bibfnamefont {C.~J.}\ \bibnamefont {Foot}}, \bibinfo {author}
  {\bibfnamefont {J.~C.}\ \bibnamefont {Frisch}}, \bibinfo {author}
  {\bibfnamefont {B.~E.}\ \bibnamefont {Garber}}, \bibinfo {author}
  {\bibfnamefont {S.}~\bibnamefont {Geer}}, \bibinfo {author} {\bibfnamefont
  {V.}~\bibnamefont {Gibson}}, \bibinfo {author} {\bibfnamefont
  {J.}~\bibnamefont {Glick}}, \bibinfo {author} {\bibfnamefont {P.~W.}\
  \bibnamefont {Graham}}, \bibinfo {author} {\bibfnamefont {S.~R.}\
  \bibnamefont {Hahn}}, \emph {et~al.},\ }\bibfield  {title} {\bibinfo {title}
  {{Matter-wave Atomic Gradiometer Interferometric Sensor (MAGIS-100)}},\
  }\href {https://doi.org/10.1088/2058-9565/abf719} {\bibfield  {journal}
  {\bibinfo  {journal} {Quantum Sci. Technol.}\ }\textbf {\bibinfo {volume}
  {6}},\ \bibinfo {pages} {044003} (\bibinfo {year} {2021})}\BibitemShut
  {NoStop}%
\bibitem [{\citenamefont {Clad\'e}\ \emph {et~al.}(2009)\citenamefont
  {Clad\'e}, \citenamefont {Guellati-Kh\'elifa}, \citenamefont {Nez},\ and\
  \citenamefont {Biraben}}]{clade:2009:large}%
  \BibitemOpen
  \bibfield  {author} {\bibinfo {author} {\bibfnamefont {P.}~\bibnamefont
  {Clad\'e}}, \bibinfo {author} {\bibfnamefont {S.}~\bibnamefont
  {Guellati-Kh\'elifa}}, \bibinfo {author} {\bibfnamefont {F.}~\bibnamefont
  {Nez}},\ and\ \bibinfo {author} {\bibfnamefont {F.}~\bibnamefont {Biraben}},\
  }\bibfield  {title} {\bibinfo {title} {{Large Momentum Beam Splitter Using
  Bloch Oscillations}},\ }\href
  {https://doi.org/10.1103/PhysRevLett.102.240402} {\bibfield  {journal}
  {\bibinfo  {journal} {Phys. Rev. Lett.}\ }\textbf {\bibinfo {volume} {102}},\
  \bibinfo {pages} {240402} (\bibinfo {year} {2009})}\BibitemShut {NoStop}%
\bibitem [{\citenamefont {McDonald}\ \emph {et~al.}(2013)\citenamefont
  {McDonald}, \citenamefont {Kuhn}, \citenamefont {Bennetts}, \citenamefont
  {Debs}, \citenamefont {Hardman}, \citenamefont {Johnsson}, \citenamefont
  {Close},\ and\ \citenamefont {Robins}}]{mcDonald:2013:80hbark}%
  \BibitemOpen
  \bibfield  {author} {\bibinfo {author} {\bibfnamefont {G.~D.}\ \bibnamefont
  {McDonald}}, \bibinfo {author} {\bibfnamefont {C.~C.~N.}\ \bibnamefont
  {Kuhn}}, \bibinfo {author} {\bibfnamefont {S.}~\bibnamefont {Bennetts}},
  \bibinfo {author} {\bibfnamefont {J.~E.}\ \bibnamefont {Debs}}, \bibinfo
  {author} {\bibfnamefont {K.~S.}\ \bibnamefont {Hardman}}, \bibinfo {author}
  {\bibfnamefont {M.}~\bibnamefont {Johnsson}}, \bibinfo {author}
  {\bibfnamefont {J.~D.}\ \bibnamefont {Close}},\ and\ \bibinfo {author}
  {\bibfnamefont {N.~P.}\ \bibnamefont {Robins}},\ }\bibfield  {title}
  {\bibinfo {title} {{$80\ensuremath{\hbar}k$ momentum separation with Bloch
  oscillations in an optically guided atom interferometer}},\ }\href
  {https://doi.org/10.1103/PhysRevA.88.053620} {\bibfield  {journal} {\bibinfo
  {journal} {Phys. Rev. A}\ }\textbf {\bibinfo {volume} {88}},\ \bibinfo
  {pages} {053620} (\bibinfo {year} {2013})}\BibitemShut {NoStop}%
\bibitem [{\citenamefont {Pagel}\ \emph {et~al.}(2020)\citenamefont {Pagel},
  \citenamefont {Zhong}, \citenamefont {Parker}, \citenamefont {Olund},
  \citenamefont {Yao} \emph {et~al.}}]{pagel:2020:symmetric}%
  \BibitemOpen
  \bibfield  {author} {\bibinfo {author} {\bibfnamefont {Z.}~\bibnamefont
  {Pagel}}, \bibinfo {author} {\bibfnamefont {W.}~\bibnamefont {Zhong}},
  \bibinfo {author} {\bibfnamefont {R.~H.}\ \bibnamefont {Parker}}, \bibinfo
  {author} {\bibfnamefont {C.~T.}\ \bibnamefont {Olund}}, \bibinfo {author}
  {\bibfnamefont {N.~Y.}\ \bibnamefont {Yao}}, \emph {et~al.},\ }\bibfield
  {title} {\bibinfo {title} {{Symmetric Bloch oscillations of matter waves}},\
  }\href {https://doi.org/10.1103/PhysRevA.102.053312} {\bibfield  {journal}
  {\bibinfo  {journal} {Phys. Rev. A}\ }\textbf {\bibinfo {volume} {102}},\
  \bibinfo {pages} {053312} (\bibinfo {year} {2020})}\BibitemShut {NoStop}%
\bibitem [{\citenamefont {Rahman}\ \emph {et~al.}(2024)\citenamefont {Rahman},
  \citenamefont {Wirth-Singh}, \citenamefont {Ivanov}, \citenamefont
  {Gochnauer}, \citenamefont {Hough} \emph {et~al.}}]{rahman:2024:bloch}%
  \BibitemOpen
  \bibfield  {author} {\bibinfo {author} {\bibfnamefont {T.}~\bibnamefont
  {Rahman}}, \bibinfo {author} {\bibfnamefont {A.}~\bibnamefont {Wirth-Singh}},
  \bibinfo {author} {\bibfnamefont {A.}~\bibnamefont {Ivanov}}, \bibinfo
  {author} {\bibfnamefont {D.}~\bibnamefont {Gochnauer}}, \bibinfo {author}
  {\bibfnamefont {E.}~\bibnamefont {Hough}}, \emph {et~al.},\ }\bibfield
  {title} {\bibinfo {title} {{Bloch oscillation phases investigated by
  multipath St\"uckelberg atom interferometry}},\ }\href
  {https://doi.org/10.1103/PhysRevResearch.6.L022012} {\bibfield  {journal}
  {\bibinfo  {journal} {Phys. Rev. Res.}\ }\textbf {\bibinfo {volume} {6}},\
  \bibinfo {pages} {L022012} (\bibinfo {year} {2024})}\BibitemShut {NoStop}%
\bibitem [{\citenamefont {Fitzek}\ \emph {et~al.}(2024)\citenamefont {Fitzek},
  \citenamefont {Kirsten-Siem\ss{}}, \citenamefont {Rasel}, \citenamefont
  {Gaaloul},\ and\ \citenamefont {Hammerer}}]{fitzek:2023:accurate}%
  \BibitemOpen
  \bibfield  {author} {\bibinfo {author} {\bibfnamefont {F.}~\bibnamefont
  {Fitzek}}, \bibinfo {author} {\bibfnamefont {J.-N.}\ \bibnamefont
  {Kirsten-Siem\ss{}}}, \bibinfo {author} {\bibfnamefont {E.~M.}\ \bibnamefont
  {Rasel}}, \bibinfo {author} {\bibfnamefont {N.}~\bibnamefont {Gaaloul}},\
  and\ \bibinfo {author} {\bibfnamefont {K.}~\bibnamefont {Hammerer}},\
  }\bibfield  {title} {\bibinfo {title} {{Accurate and efficient
  Bloch-oscillation-enhanced atom interferometry}},\ }\href
  {https://doi.org/10.1103/PhysRevResearch.6.L032028} {\bibfield  {journal}
  {\bibinfo  {journal} {Phys. Rev. Res.}\ }\textbf {\bibinfo {volume} {6}},\
  \bibinfo {pages} {L032028} (\bibinfo {year} {2024})}\BibitemShut {NoStop}%
\bibitem [{\citenamefont {Rudolph}\ \emph {et~al.}(2020)\citenamefont
  {Rudolph}, \citenamefont {Wilkason}, \citenamefont {Nantel}, \citenamefont
  {Swan}, \citenamefont {Holland} \emph {et~al.}}]{rudolph:2020:largemomentum}%
  \BibitemOpen
  \bibfield  {author} {\bibinfo {author} {\bibfnamefont {J.}~\bibnamefont
  {Rudolph}}, \bibinfo {author} {\bibfnamefont {T.}~\bibnamefont {Wilkason}},
  \bibinfo {author} {\bibfnamefont {M.}~\bibnamefont {Nantel}}, \bibinfo
  {author} {\bibfnamefont {H.}~\bibnamefont {Swan}}, \bibinfo {author}
  {\bibfnamefont {C.~M.}\ \bibnamefont {Holland}}, \emph {et~al.},\ }\bibfield
  {title} {\bibinfo {title} {{Large Momentum Transfer Clock Atom Interferometry
  on the 689 nm Intercombination Line of Strontium}},\ }\href
  {https://doi.org/10.1103/PhysRevLett.124.083604} {\bibfield  {journal}
  {\bibinfo  {journal} {Phys. Rev. Lett.}\ }\textbf {\bibinfo {volume} {124}},\
  \bibinfo {pages} {083604} (\bibinfo {year} {2020})}\BibitemShut {NoStop}%
\bibitem [{\citenamefont {Berg}\ \emph {et~al.}(2015)\citenamefont {Berg},
  \citenamefont {Abend}, \citenamefont {Tackmann}, \citenamefont {Schubert},
  \citenamefont {Giese} \emph {et~al.}}]{berg:2015:composite-light}%
  \BibitemOpen
  \bibfield  {author} {\bibinfo {author} {\bibfnamefont {P.}~\bibnamefont
  {Berg}}, \bibinfo {author} {\bibfnamefont {S.}~\bibnamefont {Abend}},
  \bibinfo {author} {\bibfnamefont {G.}~\bibnamefont {Tackmann}}, \bibinfo
  {author} {\bibfnamefont {C.}~\bibnamefont {Schubert}}, \bibinfo {author}
  {\bibfnamefont {E.}~\bibnamefont {Giese}}, \emph {et~al.},\ }\bibfield
  {title} {\bibinfo {title} {{Composite-Light-Pulse Technique for
  High-Precision Atom Interferometry}},\ }\href
  {https://doi.org/10.1103/PhysRevLett.114.063002} {\bibfield  {journal}
  {\bibinfo  {journal} {Phys. Rev. Lett.}\ }\textbf {\bibinfo {volume} {114}},\
  \bibinfo {pages} {063002} (\bibinfo {year} {2015})}\BibitemShut {NoStop}%
\bibitem [{\citenamefont {McGuirk}\ \emph {et~al.}(2000)\citenamefont
  {McGuirk}, \citenamefont {Snadden},\ and\ \citenamefont
  {Kasevich}}]{mcguirk:2000:largearea}%
  \BibitemOpen
  \bibfield  {author} {\bibinfo {author} {\bibfnamefont {J.~M.}\ \bibnamefont
  {McGuirk}}, \bibinfo {author} {\bibfnamefont {M.~J.}\ \bibnamefont
  {Snadden}},\ and\ \bibinfo {author} {\bibfnamefont {M.~A.}\ \bibnamefont
  {Kasevich}},\ }\bibfield  {title} {\bibinfo {title} {{Large Area Light-Pulse
  Atom Interferometry}},\ }\href {https://doi.org/10.1103/PhysRevLett.85.4498}
  {\bibfield  {journal} {\bibinfo  {journal} {Phys. Rev. Lett.}\ }\textbf
  {\bibinfo {volume} {85}},\ \bibinfo {pages} {4498} (\bibinfo {year}
  {2000})}\BibitemShut {NoStop}%
\bibitem [{\citenamefont {L\'ev\`eque}\ \emph {et~al.}(2009)\citenamefont
  {L\'ev\`eque}, \citenamefont {Gauguet}, \citenamefont {Michaud},
  \citenamefont {Pereira Dos~Santos},\ and\ \citenamefont
  {Landragin}}]{leveque:2009:enhancing}%
  \BibitemOpen
  \bibfield  {author} {\bibinfo {author} {\bibfnamefont {T.}~\bibnamefont
  {L\'ev\`eque}}, \bibinfo {author} {\bibfnamefont {A.}~\bibnamefont
  {Gauguet}}, \bibinfo {author} {\bibfnamefont {F.}~\bibnamefont {Michaud}},
  \bibinfo {author} {\bibfnamefont {F.}~\bibnamefont {Pereira Dos~Santos}},\
  and\ \bibinfo {author} {\bibfnamefont {A.}~\bibnamefont {Landragin}},\
  }\bibfield  {title} {\bibinfo {title} {{Enhancing the Area of a Raman Atom
  Interferometer Using a Versatile Double-Diffraction Technique}},\ }\href
  {https://doi.org/10.1103/PhysRevLett.103.080405} {\bibfield  {journal}
  {\bibinfo  {journal} {Phys. Rev. Lett.}\ }\textbf {\bibinfo {volume} {103}},\
  \bibinfo {pages} {080405} (\bibinfo {year} {2009})}\BibitemShut {NoStop}%
\bibitem [{\citenamefont {Giese}\ \emph {et~al.}(2013)\citenamefont {Giese},
  \citenamefont {Roura}, \citenamefont {Tackmann}, \citenamefont {Rasel},\ and\
  \citenamefont {Schleich}}]{giese:2013:double}%
  \BibitemOpen
  \bibfield  {author} {\bibinfo {author} {\bibfnamefont {E.}~\bibnamefont
  {Giese}}, \bibinfo {author} {\bibfnamefont {A.}~\bibnamefont {Roura}},
  \bibinfo {author} {\bibfnamefont {G.}~\bibnamefont {Tackmann}}, \bibinfo
  {author} {\bibfnamefont {E.~M.}\ \bibnamefont {Rasel}},\ and\ \bibinfo
  {author} {\bibfnamefont {W.~P.}\ \bibnamefont {Schleich}},\ }\bibfield
  {title} {\bibinfo {title} {{Double Bragg diffraction: A tool for atom
  optics}},\ }\href {https://doi.org/10.1103/PhysRevA.88.053608} {\bibfield
  {journal} {\bibinfo  {journal} {Phys. Rev. A}\ }\textbf {\bibinfo {volume}
  {88}},\ \bibinfo {pages} {053608} (\bibinfo {year} {2013})}\BibitemShut
  {NoStop}%
\bibitem [{\citenamefont {K{\"u}ber}\ \emph {et~al.}(2016)\citenamefont
  {K{\"u}ber}, \citenamefont {Schmaltz},\ and\ \citenamefont
  {Birkl}}]{kueber:2016:experimental}%
  \BibitemOpen
  \bibfield  {author} {\bibinfo {author} {\bibfnamefont {J.}~\bibnamefont
  {K{\"u}ber}}, \bibinfo {author} {\bibfnamefont {F.}~\bibnamefont
  {Schmaltz}},\ and\ \bibinfo {author} {\bibfnamefont {G.}~\bibnamefont
  {Birkl}},\ }\href {https://doi.org/10.48550/arXiv.1603.08826} {\bibinfo
  {title} {{Experimental realization of double Bragg diffraction: robust
  beamsplitters, mirrors, and interferometers for Bose-Einstein condensates}}}
  (\bibinfo {year} {2016}),\ \Eprint {https://arxiv.org/abs/1603.08826}
  {arXiv:1603.08826 [cond-mat.quant-gas]} \BibitemShut {NoStop}%
\bibitem [{\citenamefont {Hartmann}\ \emph
  {et~al.}(2020{\natexlab{a}})\citenamefont {Hartmann}, \citenamefont
  {Jenewein}, \citenamefont {Giese}, \citenamefont {Abend}, \citenamefont
  {Roura} \emph {et~al.}}]{hartmann:2020:regimes}%
  \BibitemOpen
  \bibfield  {author} {\bibinfo {author} {\bibfnamefont {S.}~\bibnamefont
  {Hartmann}}, \bibinfo {author} {\bibfnamefont {J.}~\bibnamefont {Jenewein}},
  \bibinfo {author} {\bibfnamefont {E.}~\bibnamefont {Giese}}, \bibinfo
  {author} {\bibfnamefont {S.}~\bibnamefont {Abend}}, \bibinfo {author}
  {\bibfnamefont {A.}~\bibnamefont {Roura}}, \emph {et~al.},\ }\bibfield
  {title} {\bibinfo {title} {{Regimes of atomic diffraction: Raman versus Bragg
  diffraction in retroreflective geometries}},\ }\href
  {https://doi.org/10.1103/PhysRevA.101.053610} {\bibfield  {journal} {\bibinfo
   {journal} {Phys. Rev. A}\ }\textbf {\bibinfo {volume} {101}},\ \bibinfo
  {pages} {053610} (\bibinfo {year} {2020}{\natexlab{a}})}\BibitemShut
  {NoStop}%
\bibitem [{\citenamefont {Hartmann}\ \emph
  {et~al.}(2020{\natexlab{b}})\citenamefont {Hartmann}, \citenamefont
  {Jenewein}, \citenamefont {Abend}, \citenamefont {Roura},\ and\ \citenamefont
  {Giese}}]{hartmann:2020:atomicraman}%
  \BibitemOpen
  \bibfield  {author} {\bibinfo {author} {\bibfnamefont {S.}~\bibnamefont
  {Hartmann}}, \bibinfo {author} {\bibfnamefont {J.}~\bibnamefont {Jenewein}},
  \bibinfo {author} {\bibfnamefont {S.}~\bibnamefont {Abend}}, \bibinfo
  {author} {\bibfnamefont {A.}~\bibnamefont {Roura}},\ and\ \bibinfo {author}
  {\bibfnamefont {E.}~\bibnamefont {Giese}},\ }\bibfield  {title} {\bibinfo
  {title} {{Atomic Raman scattering: Third-order diffraction in a double
  geometry}},\ }\href {https://doi.org/10.1103/PhysRevA.102.063326} {\bibfield
  {journal} {\bibinfo  {journal} {Phys. Rev. A}\ }\textbf {\bibinfo {volume}
  {102}},\ \bibinfo {pages} {063326} (\bibinfo {year}
  {2020}{\natexlab{b}})}\BibitemShut {NoStop}%
\bibitem [{\citenamefont {M\"uller}\ \emph
  {et~al.}(2008{\natexlab{a}})\citenamefont {M\"uller}, \citenamefont {Chiow},
  \citenamefont {Long}, \citenamefont {Herrmann},\ and\ \citenamefont
  {Chu}}]{mueller:2008:atom}%
  \BibitemOpen
  \bibfield  {author} {\bibinfo {author} {\bibfnamefont {H.}~\bibnamefont
  {M\"uller}}, \bibinfo {author} {\bibfnamefont {S.-w.}\ \bibnamefont {Chiow}},
  \bibinfo {author} {\bibfnamefont {Q.}~\bibnamefont {Long}}, \bibinfo {author}
  {\bibfnamefont {S.}~\bibnamefont {Herrmann}},\ and\ \bibinfo {author}
  {\bibfnamefont {S.}~\bibnamefont {Chu}},\ }\bibfield  {title} {\bibinfo
  {title} {{Atom Interferometry with up to 24-Photon-Momentum-Transfer Beam
  Splitters}},\ }\href {https://doi.org/10.1103/PhysRevLett.100.180405}
  {\bibfield  {journal} {\bibinfo  {journal} {Phys. Rev. Lett.}\ }\textbf
  {\bibinfo {volume} {100}},\ \bibinfo {pages} {180405} (\bibinfo {year}
  {2008}{\natexlab{a}})}\BibitemShut {NoStop}%
\bibitem [{\citenamefont {Ahlers}\ \emph {et~al.}(2016)\citenamefont {Ahlers},
  \citenamefont {M\"untinga}, \citenamefont {Wenzlawski}, \citenamefont
  {Krutzik}, \citenamefont {Tackmann}, \citenamefont {Abend}, \citenamefont
  {Gaaloul}, \citenamefont {Giese}, \citenamefont {Roura}, \citenamefont
  {Kuhl}, \citenamefont {L\"ammerzahl}, \citenamefont {Peters}, \citenamefont
  {Windpassinger}, \citenamefont {Sengstock}, \citenamefont {Schleich},
  \citenamefont {Ertmer},\ and\ \citenamefont
  {Rasel}}]{ahlers:2016:doublebragg}%
  \BibitemOpen
  \bibfield  {author} {\bibinfo {author} {\bibfnamefont {H.}~\bibnamefont
  {Ahlers}}, \bibinfo {author} {\bibfnamefont {H.}~\bibnamefont {M\"untinga}},
  \bibinfo {author} {\bibfnamefont {A.}~\bibnamefont {Wenzlawski}}, \bibinfo
  {author} {\bibfnamefont {M.}~\bibnamefont {Krutzik}}, \bibinfo {author}
  {\bibfnamefont {G.}~\bibnamefont {Tackmann}}, \bibinfo {author}
  {\bibfnamefont {S.}~\bibnamefont {Abend}}, \bibinfo {author} {\bibfnamefont
  {N.}~\bibnamefont {Gaaloul}}, \bibinfo {author} {\bibfnamefont
  {E.}~\bibnamefont {Giese}}, \bibinfo {author} {\bibfnamefont
  {A.}~\bibnamefont {Roura}}, \bibinfo {author} {\bibfnamefont
  {R.}~\bibnamefont {Kuhl}}, \bibinfo {author} {\bibfnamefont {C.}~\bibnamefont
  {L\"ammerzahl}}, \bibinfo {author} {\bibfnamefont {A.}~\bibnamefont
  {Peters}}, \bibinfo {author} {\bibfnamefont {P.}~\bibnamefont
  {Windpassinger}}, \bibinfo {author} {\bibfnamefont {K.}~\bibnamefont
  {Sengstock}}, \bibinfo {author} {\bibfnamefont {W.~P.}\ \bibnamefont
  {Schleich}}, \bibinfo {author} {\bibfnamefont {W.}~\bibnamefont {Ertmer}},\
  and\ \bibinfo {author} {\bibfnamefont {E.~M.}\ \bibnamefont {Rasel}},\
  }\bibfield  {title} {\bibinfo {title} {{Double Bragg Interferometry}},\
  }\href {https://doi.org/10.1103/PhysRevLett.116.173601} {\bibfield  {journal}
  {\bibinfo  {journal} {Phys. Rev. Lett.}\ }\textbf {\bibinfo {volume} {116}},\
  \bibinfo {pages} {173601} (\bibinfo {year} {2016})}\BibitemShut {NoStop}%
\bibitem [{\citenamefont {M\"uller}\ \emph {et~al.}(2009)\citenamefont
  {M\"uller}, \citenamefont {Chiow}, \citenamefont {Herrmann},\ and\
  \citenamefont {Chu}}]{mueller:2009:atom}%
  \BibitemOpen
  \bibfield  {author} {\bibinfo {author} {\bibfnamefont {H.}~\bibnamefont
  {M\"uller}}, \bibinfo {author} {\bibfnamefont {S.-w.}\ \bibnamefont {Chiow}},
  \bibinfo {author} {\bibfnamefont {S.}~\bibnamefont {Herrmann}},\ and\
  \bibinfo {author} {\bibfnamefont {S.}~\bibnamefont {Chu}},\ }\bibfield
  {title} {\bibinfo {title} {{Atom Interferometers with Scalable Enclosed
  Area}},\ }\href {https://doi.org/10.1103/PhysRevLett.102.240403} {\bibfield
  {journal} {\bibinfo  {journal} {Phys. Rev. Lett.}\ }\textbf {\bibinfo
  {volume} {102}},\ \bibinfo {pages} {240403} (\bibinfo {year}
  {2009})}\BibitemShut {NoStop}%
\bibitem [{\citenamefont {Chiarotti}\ \emph {et~al.}(2022)\citenamefont
  {Chiarotti}, \citenamefont {Tinsley}, \citenamefont {Bandarupally},
  \citenamefont {Manzoor}, \citenamefont {Sacco} \emph
  {et~al.}}]{chiarotti:2022:prx}%
  \BibitemOpen
  \bibfield  {author} {\bibinfo {author} {\bibfnamefont {M.}~\bibnamefont
  {Chiarotti}}, \bibinfo {author} {\bibfnamefont {J.~N.}\ \bibnamefont
  {Tinsley}}, \bibinfo {author} {\bibfnamefont {S.}~\bibnamefont
  {Bandarupally}}, \bibinfo {author} {\bibfnamefont {S.}~\bibnamefont
  {Manzoor}}, \bibinfo {author} {\bibfnamefont {M.}~\bibnamefont {Sacco}},
  \emph {et~al.},\ }\bibfield  {title} {\bibinfo {title} {{Practical Limits for
  Large-Momentum-Transfer Clock Atom Interferometers}},\ }\href
  {https://doi.org/10.1103/PRXQuantum.3.030348} {\bibfield  {journal} {\bibinfo
   {journal} {PRX Quantum}\ }\textbf {\bibinfo {volume} {3}},\ \bibinfo {pages}
  {030348} (\bibinfo {year} {2022})}\BibitemShut {NoStop}%
\bibitem [{\citenamefont {B\"uchner}\ \emph {et~al.}(2003)\citenamefont
  {B\"uchner}, \citenamefont {Delhuille}, \citenamefont {Miffre}, \citenamefont
  {Robilliard}, \citenamefont {Vigu\'e} \emph
  {et~al.}}]{buechner:2003:diffraction}%
  \BibitemOpen
  \bibfield  {author} {\bibinfo {author} {\bibfnamefont {M.}~\bibnamefont
  {B\"uchner}}, \bibinfo {author} {\bibfnamefont {R.}~\bibnamefont
  {Delhuille}}, \bibinfo {author} {\bibfnamefont {A.}~\bibnamefont {Miffre}},
  \bibinfo {author} {\bibfnamefont {C.}~\bibnamefont {Robilliard}}, \bibinfo
  {author} {\bibfnamefont {J.}~\bibnamefont {Vigu\'e}}, \emph {et~al.},\
  }\bibfield  {title} {\bibinfo {title} {{Diffraction phases in atom
  interferometers}},\ }\href {https://doi.org/10.1103/PhysRevA.68.013607}
  {\bibfield  {journal} {\bibinfo  {journal} {Phys. Rev. A}\ }\textbf {\bibinfo
  {volume} {68}},\ \bibinfo {pages} {013607} (\bibinfo {year}
  {2003})}\BibitemShut {NoStop}%
\bibitem [{\citenamefont {Estey}\ \emph {et~al.}(2015)\citenamefont {Estey},
  \citenamefont {Yu}, \citenamefont {M\"uller}, \citenamefont {Kuan},\ and\
  \citenamefont {Lan}}]{estey:2015:high-resolution}%
  \BibitemOpen
  \bibfield  {author} {\bibinfo {author} {\bibfnamefont {B.}~\bibnamefont
  {Estey}}, \bibinfo {author} {\bibfnamefont {C.}~\bibnamefont {Yu}}, \bibinfo
  {author} {\bibfnamefont {H.}~\bibnamefont {M\"uller}}, \bibinfo {author}
  {\bibfnamefont {P.-C.}\ \bibnamefont {Kuan}},\ and\ \bibinfo {author}
  {\bibfnamefont {S.-Y.}\ \bibnamefont {Lan}},\ }\bibfield  {title} {\bibinfo
  {title} {{High-Resolution Atom Interferometers with Suppressed Diffraction
  Phases}},\ }\href {https://doi.org/10.1103/PhysRevLett.115.083002} {\bibfield
   {journal} {\bibinfo  {journal} {Phys. Rev. Lett.}\ }\textbf {\bibinfo
  {volume} {115}},\ \bibinfo {pages} {083002} (\bibinfo {year}
  {2015})}\BibitemShut {NoStop}%
\bibitem [{\citenamefont {B\'eguin}\ \emph {et~al.}(2022)\citenamefont
  {B\'eguin}, \citenamefont {Rodzinka}, \citenamefont {Vigu\'e}, \citenamefont
  {Allard},\ and\ \citenamefont {Gauguet}}]{beguin:2022:characterization}%
  \BibitemOpen
  \bibfield  {author} {\bibinfo {author} {\bibfnamefont {A.}~\bibnamefont
  {B\'eguin}}, \bibinfo {author} {\bibfnamefont {T.}~\bibnamefont {Rodzinka}},
  \bibinfo {author} {\bibfnamefont {J.}~\bibnamefont {Vigu\'e}}, \bibinfo
  {author} {\bibfnamefont {B.}~\bibnamefont {Allard}},\ and\ \bibinfo {author}
  {\bibfnamefont {A.}~\bibnamefont {Gauguet}},\ }\bibfield  {title} {\bibinfo
  {title} {{Characterization of an atom interferometer in the quasi-Bragg
  regime}},\ }\href {https://doi.org/10.1103/PhysRevA.105.033302} {\bibfield
  {journal} {\bibinfo  {journal} {Phys. Rev. A}\ }\textbf {\bibinfo {volume}
  {105}},\ \bibinfo {pages} {033302} (\bibinfo {year} {2022})}\BibitemShut
  {NoStop}%
\bibitem [{\citenamefont {Altin}\ \emph {et~al.}(2013)\citenamefont {Altin},
  \citenamefont {Johnsson}, \citenamefont {Negnevitsky}, \citenamefont
  {Dennis}, \citenamefont {Anderson}, \citenamefont {Debs}, \citenamefont
  {Szigeti}, \citenamefont {Hardman}, \citenamefont {Bennetts}, \citenamefont
  {McDonald}, \citenamefont {Turner}, \citenamefont {Close},\ and\
  \citenamefont {Robins}}]{Altin:2013:precision}%
  \BibitemOpen
  \bibfield  {author} {\bibinfo {author} {\bibfnamefont {P.~A.}\ \bibnamefont
  {Altin}}, \bibinfo {author} {\bibfnamefont {M.~T.}\ \bibnamefont {Johnsson}},
  \bibinfo {author} {\bibfnamefont {V.}~\bibnamefont {Negnevitsky}}, \bibinfo
  {author} {\bibfnamefont {G.~R.}\ \bibnamefont {Dennis}}, \bibinfo {author}
  {\bibfnamefont {R.~P.}\ \bibnamefont {Anderson}}, \bibinfo {author}
  {\bibfnamefont {J.~E.}\ \bibnamefont {Debs}}, \bibinfo {author}
  {\bibfnamefont {S.~S.}\ \bibnamefont {Szigeti}}, \bibinfo {author}
  {\bibfnamefont {K.~S.}\ \bibnamefont {Hardman}}, \bibinfo {author}
  {\bibfnamefont {S.}~\bibnamefont {Bennetts}}, \bibinfo {author}
  {\bibfnamefont {G.~D.}\ \bibnamefont {McDonald}}, \bibinfo {author}
  {\bibfnamefont {L.~D.}\ \bibnamefont {Turner}}, \bibinfo {author}
  {\bibfnamefont {J.~D.}\ \bibnamefont {Close}},\ and\ \bibinfo {author}
  {\bibfnamefont {N.~P.}\ \bibnamefont {Robins}},\ }\bibfield  {title}
  {\bibinfo {title} {{Precision atomic gravimeter based on Bragg
  diffraction}},\ }\href {https://doi.org/10.1088/1367-2630/15/2/023009}
  {\bibfield  {journal} {\bibinfo  {journal} {New J. Phys.}\ }\textbf {\bibinfo
  {volume} {15}},\ \bibinfo {pages} {023009} (\bibinfo {year}
  {2013})}\BibitemShut {NoStop}%
\bibitem [{\citenamefont {Parker}\ \emph {et~al.}(2016)\citenamefont {Parker},
  \citenamefont {Yu}, \citenamefont {Estey}, \citenamefont {Zhong},
  \citenamefont {Huang},\ and\ \citenamefont
  {M\"uller}}]{parker:2016:controlling}%
  \BibitemOpen
  \bibfield  {author} {\bibinfo {author} {\bibfnamefont {R.~H.}\ \bibnamefont
  {Parker}}, \bibinfo {author} {\bibfnamefont {C.}~\bibnamefont {Yu}}, \bibinfo
  {author} {\bibfnamefont {B.}~\bibnamefont {Estey}}, \bibinfo {author}
  {\bibfnamefont {W.}~\bibnamefont {Zhong}}, \bibinfo {author} {\bibfnamefont
  {E.}~\bibnamefont {Huang}},\ and\ \bibinfo {author} {\bibfnamefont
  {H.}~\bibnamefont {M\"uller}},\ }\bibfield  {title} {\bibinfo {title}
  {{Controlling the multiport nature of Bragg diffraction in atom
  interferometry}},\ }\href {https://doi.org/10.1103/PhysRevA.94.053618}
  {\bibfield  {journal} {\bibinfo  {journal} {Phys. Rev. A}\ }\textbf {\bibinfo
  {volume} {94}},\ \bibinfo {pages} {053618} (\bibinfo {year}
  {2016})}\BibitemShut {NoStop}%
\bibitem [{\citenamefont {Jenewein}\ \emph {et~al.}(2022)\citenamefont
  {Jenewein}, \citenamefont {Hartmann}, \citenamefont {Roura},\ and\
  \citenamefont {Giese}}]{jenewein:2022:bragg}%
  \BibitemOpen
  \bibfield  {author} {\bibinfo {author} {\bibfnamefont {J.}~\bibnamefont
  {Jenewein}}, \bibinfo {author} {\bibfnamefont {S.}~\bibnamefont {Hartmann}},
  \bibinfo {author} {\bibfnamefont {A.}~\bibnamefont {Roura}},\ and\ \bibinfo
  {author} {\bibfnamefont {E.}~\bibnamefont {Giese}},\ }\bibfield  {title}
  {\bibinfo {title} {{Bragg-diffraction-induced imperfections of the signal in
  retroreflective atom interferometers}},\ }\href
  {https://doi.org/10.1103/PhysRevA.105.063316} {\bibfield  {journal} {\bibinfo
   {journal} {Phys. Rev. A}\ }\textbf {\bibinfo {volume} {105}},\ \bibinfo
  {pages} {063316} (\bibinfo {year} {2022})}\BibitemShut {NoStop}%
\bibitem [{\citenamefont {Wilkason}\ \emph {et~al.}(2022)\citenamefont
  {Wilkason}, \citenamefont {Nantel}, \citenamefont {Rudolph}, \citenamefont
  {Jiang}, \citenamefont {Garber} \emph
  {et~al.}}]{wilkason:2022:atominterferometry}%
  \BibitemOpen
  \bibfield  {author} {\bibinfo {author} {\bibfnamefont {T.}~\bibnamefont
  {Wilkason}}, \bibinfo {author} {\bibfnamefont {M.}~\bibnamefont {Nantel}},
  \bibinfo {author} {\bibfnamefont {J.}~\bibnamefont {Rudolph}}, \bibinfo
  {author} {\bibfnamefont {Y.}~\bibnamefont {Jiang}}, \bibinfo {author}
  {\bibfnamefont {B.~E.}\ \bibnamefont {Garber}}, \emph {et~al.},\ }\bibfield
  {title} {\bibinfo {title} {{Atom Interferometry with Floquet Atom Optics}},\
  }\href {https://doi.org/10.1103/PhysRevLett.129.183202} {\bibfield  {journal}
  {\bibinfo  {journal} {Phys. Rev. Lett.}\ }\textbf {\bibinfo {volume} {129}},\
  \bibinfo {pages} {183202} (\bibinfo {year} {2022})}\BibitemShut {NoStop}%
\bibitem [{\citenamefont {Rodzinka}\ \emph {et~al.}(2024)\citenamefont
  {Rodzinka}, \citenamefont {Dionis}, \citenamefont {Calmels}, \citenamefont
  {Beldjoudi}, \citenamefont {B{\'e}guin}, \citenamefont {Gu{\'e}ry-Odelin},
  \citenamefont {Allard}, \citenamefont {Sugny},\ and\ \citenamefont
  {Gauguet}}]{rodzinka:2024:optimalfloquetengineeringlarge}%
  \BibitemOpen
  \bibfield  {author} {\bibinfo {author} {\bibfnamefont {T.}~\bibnamefont
  {Rodzinka}}, \bibinfo {author} {\bibfnamefont {E.}~\bibnamefont {Dionis}},
  \bibinfo {author} {\bibfnamefont {L.}~\bibnamefont {Calmels}}, \bibinfo
  {author} {\bibfnamefont {S.}~\bibnamefont {Beldjoudi}}, \bibinfo {author}
  {\bibfnamefont {A.}~\bibnamefont {B{\'e}guin}}, \bibinfo {author}
  {\bibfnamefont {D.}~\bibnamefont {Gu{\'e}ry-Odelin}}, \bibinfo {author}
  {\bibfnamefont {B.}~\bibnamefont {Allard}}, \bibinfo {author} {\bibfnamefont
  {D.}~\bibnamefont {Sugny}},\ and\ \bibinfo {author} {\bibfnamefont
  {A.}~\bibnamefont {Gauguet}},\ }\bibfield  {title} {\bibinfo {title}
  {{Optimal Floquet state engineering for large scale atom interferometers}},\
  }\href {https://doi.org/10.1038/s41467-024-54539-w} {\bibfield  {journal}
  {\bibinfo  {journal} {Nat. Commun.}\ }\textbf {\bibinfo {volume} {15}},\
  \bibinfo {pages} {10281} (\bibinfo {year} {2024})}\BibitemShut {NoStop}%
\bibitem [{\citenamefont {Chen}\ \emph {et~al.}(2023)\citenamefont {Chen},
  \citenamefont {Louie}, \citenamefont {Wang}, \citenamefont {Deshpande},\ and\
  \citenamefont {Kovachy}}]{chen:2023:enhancing}%
  \BibitemOpen
  \bibfield  {author} {\bibinfo {author} {\bibfnamefont {Z.}~\bibnamefont
  {Chen}}, \bibinfo {author} {\bibfnamefont {G.}~\bibnamefont {Louie}},
  \bibinfo {author} {\bibfnamefont {Y.}~\bibnamefont {Wang}}, \bibinfo {author}
  {\bibfnamefont {T.}~\bibnamefont {Deshpande}},\ and\ \bibinfo {author}
  {\bibfnamefont {T.}~\bibnamefont {Kovachy}},\ }\bibfield  {title} {\bibinfo
  {title} {{Enhancing strontium clock atom interferometry using quantum optimal
  control}},\ }\href {https://doi.org/10.1103/PhysRevA.107.063302} {\bibfield
  {journal} {\bibinfo  {journal} {Phys. Rev. A}\ }\textbf {\bibinfo {volume}
  {107}},\ \bibinfo {pages} {063302} (\bibinfo {year} {2023})}\BibitemShut
  {NoStop}%
\bibitem [{\citenamefont {Saywell}\ \emph
  {et~al.}(2020{\natexlab{a}})\citenamefont {Saywell}, \citenamefont {Carey},
  \citenamefont {Belal}, \citenamefont {Kuprov},\ and\ \citenamefont
  {Freegarde}}]{Saywell:2020:optimalcontrol}%
  \BibitemOpen
  \bibfield  {author} {\bibinfo {author} {\bibfnamefont {J.}~\bibnamefont
  {Saywell}}, \bibinfo {author} {\bibfnamefont {M.}~\bibnamefont {Carey}},
  \bibinfo {author} {\bibfnamefont {M.}~\bibnamefont {Belal}}, \bibinfo
  {author} {\bibfnamefont {I.}~\bibnamefont {Kuprov}},\ and\ \bibinfo {author}
  {\bibfnamefont {T.}~\bibnamefont {Freegarde}},\ }\bibfield  {title} {\bibinfo
  {title} {{Optimal control of Raman pulse sequences for atom
  interferometry}},\ }\href {https://doi.org/10.1088/1361-6455/ab6df6}
  {\bibfield  {journal} {\bibinfo  {journal} {J. Phys. B: At., Mol. Opt.
  Phys.}\ }\textbf {\bibinfo {volume} {53}},\ \bibinfo {pages} {085006}
  (\bibinfo {year} {2020}{\natexlab{a}})}\BibitemShut {NoStop}%
\bibitem [{\citenamefont {Saywell}\ \emph
  {et~al.}(2020{\natexlab{b}})\citenamefont {Saywell}, \citenamefont {Carey},
  \citenamefont {Kuprov},\ and\ \citenamefont
  {Freegarde}}]{saywell:2020:biselective}%
  \BibitemOpen
  \bibfield  {author} {\bibinfo {author} {\bibfnamefont {J.}~\bibnamefont
  {Saywell}}, \bibinfo {author} {\bibfnamefont {M.}~\bibnamefont {Carey}},
  \bibinfo {author} {\bibfnamefont {I.}~\bibnamefont {Kuprov}},\ and\ \bibinfo
  {author} {\bibfnamefont {T.}~\bibnamefont {Freegarde}},\ }\bibfield  {title}
  {\bibinfo {title} {{Biselective pulses for large-area atom interferometry}},\
  }\href {https://doi.org/10.1103/PhysRevA.101.063625} {\bibfield  {journal}
  {\bibinfo  {journal} {Phys. Rev. A}\ }\textbf {\bibinfo {volume} {101}},\
  \bibinfo {pages} {063625} (\bibinfo {year} {2020}{\natexlab{b}})}\BibitemShut
  {NoStop}%
\bibitem [{\citenamefont {Li}\ \emph {et~al.}(2024)\citenamefont {Li},
  \citenamefont {Mart\'{\i}nez-Lahuerta}, \citenamefont {Seckmeyer},
  \citenamefont {Hammerer},\ and\ \citenamefont {Gaaloul}}]{li:2024:robust}%
  \BibitemOpen
  \bibfield  {author} {\bibinfo {author} {\bibfnamefont {R.}~\bibnamefont
  {Li}}, \bibinfo {author} {\bibfnamefont {V.~J.}\ \bibnamefont
  {Mart\'{\i}nez-Lahuerta}}, \bibinfo {author} {\bibfnamefont {S.}~\bibnamefont
  {Seckmeyer}}, \bibinfo {author} {\bibfnamefont {K.}~\bibnamefont
  {Hammerer}},\ and\ \bibinfo {author} {\bibfnamefont {N.}~\bibnamefont
  {Gaaloul}},\ }\bibfield  {title} {\bibinfo {title} {{Robust double Bragg
  diffraction via detuning control}},\ }\href
  {https://doi.org/10.1103/PhysRevResearch.6.043236} {\bibfield  {journal}
  {\bibinfo  {journal} {Phys. Rev. Res.}\ }\textbf {\bibinfo {volume} {6}},\
  \bibinfo {pages} {043236} (\bibinfo {year} {2024})}\BibitemShut {NoStop}%
\bibitem [{\citenamefont {Louie}\ \emph {et~al.}(2023)\citenamefont {Louie},
  \citenamefont {Chen}, \citenamefont {Deshpande},\ and\ \citenamefont
  {Kovachy}}]{louie:2023:robust}%
  \BibitemOpen
  \bibfield  {author} {\bibinfo {author} {\bibfnamefont {G.}~\bibnamefont
  {Louie}}, \bibinfo {author} {\bibfnamefont {Z.}~\bibnamefont {Chen}},
  \bibinfo {author} {\bibfnamefont {T.}~\bibnamefont {Deshpande}},\ and\
  \bibinfo {author} {\bibfnamefont {T.}~\bibnamefont {Kovachy}},\ }\bibfield
  {title} {\bibinfo {title} {{Robust atom optics for Bragg atom
  interferometry}},\ }\href {https://doi.org/10.1088/1367-2630/aceb15}
  {\bibfield  {journal} {\bibinfo  {journal} {New J. Phys.}\ }\textbf {\bibinfo
  {volume} {25}},\ \bibinfo {pages} {083017} (\bibinfo {year}
  {2023})}\BibitemShut {NoStop}%
\bibitem [{\citenamefont {Saywell}\ \emph {et~al.}(2023)\citenamefont
  {Saywell}, \citenamefont {Carey}, \citenamefont {Light}, \citenamefont
  {Szigeti}, \citenamefont {Milne}, \citenamefont {Gill}, \citenamefont {Goh},
  \citenamefont {Perunicic}, \citenamefont {Wilson}, \citenamefont {Macrae},
  \citenamefont {Rischka}, \citenamefont {Everitt}, \citenamefont {Robins},
  \citenamefont {Anderson}, \citenamefont {Hush},\ and\ \citenamefont
  {Biercuk}}]{saywell:2023:enhancing}%
  \BibitemOpen
  \bibfield  {author} {\bibinfo {author} {\bibfnamefont {J.~C.}\ \bibnamefont
  {Saywell}}, \bibinfo {author} {\bibfnamefont {M.~S.}\ \bibnamefont {Carey}},
  \bibinfo {author} {\bibfnamefont {P.~S.}\ \bibnamefont {Light}}, \bibinfo
  {author} {\bibfnamefont {S.~S.}\ \bibnamefont {Szigeti}}, \bibinfo {author}
  {\bibfnamefont {A.~R.}\ \bibnamefont {Milne}}, \bibinfo {author}
  {\bibfnamefont {K.~S.}\ \bibnamefont {Gill}}, \bibinfo {author}
  {\bibfnamefont {M.~L.}\ \bibnamefont {Goh}}, \bibinfo {author} {\bibfnamefont
  {V.~S.}\ \bibnamefont {Perunicic}}, \bibinfo {author} {\bibfnamefont {N.~M.}\
  \bibnamefont {Wilson}}, \bibinfo {author} {\bibfnamefont {C.~D.}\
  \bibnamefont {Macrae}}, \bibinfo {author} {\bibfnamefont {A.}~\bibnamefont
  {Rischka}}, \bibinfo {author} {\bibfnamefont {P.~J.}\ \bibnamefont
  {Everitt}}, \bibinfo {author} {\bibfnamefont {N.~P.}\ \bibnamefont {Robins}},
  \bibinfo {author} {\bibfnamefont {R.~P.}\ \bibnamefont {Anderson}}, \bibinfo
  {author} {\bibfnamefont {M.~R.}\ \bibnamefont {Hush}},\ and\ \bibinfo
  {author} {\bibfnamefont {M.~J.}\ \bibnamefont {Biercuk}},\ }\bibfield
  {title} {\bibinfo {title} {{Enhancing the sensitivity of atom-interferometric
  inertial sensors using robust control}},\ }\href
  {https://doi.org/10.1038/s41467-023-43374-0} {\bibfield  {journal} {\bibinfo
  {journal} {Nat. Commun.}\ }\textbf {\bibinfo {volume} {14}},\ \bibinfo
  {pages} {7626} (\bibinfo {year} {2023})}\BibitemShut {NoStop}%
\bibitem [{\citenamefont {Wang}\ \emph {et~al.}(2024)\citenamefont {Wang},
  \citenamefont {Glick}, \citenamefont {Deshpande}, \citenamefont {DeRose},
  \citenamefont {Saraf}, \citenamefont {Sachdeva}, \citenamefont {Jiang},
  \citenamefont {Chen},\ and\ \citenamefont {Kovachy}}]{wang:2024:arxiv}%
  \BibitemOpen
  \bibfield  {author} {\bibinfo {author} {\bibfnamefont {Y.}~\bibnamefont
  {Wang}}, \bibinfo {author} {\bibfnamefont {J.}~\bibnamefont {Glick}},
  \bibinfo {author} {\bibfnamefont {T.}~\bibnamefont {Deshpande}}, \bibinfo
  {author} {\bibfnamefont {K.}~\bibnamefont {DeRose}}, \bibinfo {author}
  {\bibfnamefont {S.}~\bibnamefont {Saraf}}, \bibinfo {author} {\bibfnamefont
  {N.}~\bibnamefont {Sachdeva}}, \bibinfo {author} {\bibfnamefont
  {K.}~\bibnamefont {Jiang}}, \bibinfo {author} {\bibfnamefont
  {Z.}~\bibnamefont {Chen}},\ and\ \bibinfo {author} {\bibfnamefont
  {T.}~\bibnamefont {Kovachy}},\ }\bibfield  {title} {\bibinfo {title} {Robust
  quantum control via multipath interference for thousandfold phase
  amplification in a resonant atom interferometer},\ }\href
  {https://doi.org/10.1103/PhysRevLett.133.243403} {\bibfield  {journal}
  {\bibinfo  {journal} {Phys. Rev. Lett.}\ }\textbf {\bibinfo {volume} {133}},\
  \bibinfo {pages} {243403} (\bibinfo {year} {2024})}\BibitemShut {NoStop}%
\bibitem [{\citenamefont {B\'eguin}\ \emph {et~al.}(2023)\citenamefont
  {B\'eguin}, \citenamefont {Rodzinka}, \citenamefont {Calmels}, \citenamefont
  {Allard},\ and\ \citenamefont {Gauguet}}]{beguin:2023:atominterferometry}%
  \BibitemOpen
  \bibfield  {author} {\bibinfo {author} {\bibfnamefont {A.}~\bibnamefont
  {B\'eguin}}, \bibinfo {author} {\bibfnamefont {T.}~\bibnamefont {Rodzinka}},
  \bibinfo {author} {\bibfnamefont {L.}~\bibnamefont {Calmels}}, \bibinfo
  {author} {\bibfnamefont {B.}~\bibnamefont {Allard}},\ and\ \bibinfo {author}
  {\bibfnamefont {A.}~\bibnamefont {Gauguet}},\ }\bibfield  {title} {\bibinfo
  {title} {{Atom Interferometry with Coherent Enhancement of Bragg Pulse
  Sequences}},\ }\href {https://doi.org/10.1103/PhysRevLett.131.143401}
  {\bibfield  {journal} {\bibinfo  {journal} {Phys. Rev. Lett.}\ }\textbf
  {\bibinfo {volume} {131}},\ \bibinfo {pages} {143401} (\bibinfo {year}
  {2023})}\BibitemShut {NoStop}%
\bibitem [{\citenamefont {Kirsten-Siem\ss{}}\ \emph {et~al.}(2023)\citenamefont
  {Kirsten-Siem\ss{}}, \citenamefont {Fitzek}, \citenamefont {Schubert},
  \citenamefont {Rasel}, \citenamefont {Gaaloul},\ and\ \citenamefont
  {Hammerer}}]{kirsten:2023:prl}%
  \BibitemOpen
  \bibfield  {author} {\bibinfo {author} {\bibfnamefont {J.-N.}\ \bibnamefont
  {Kirsten-Siem\ss{}}}, \bibinfo {author} {\bibfnamefont {F.}~\bibnamefont
  {Fitzek}}, \bibinfo {author} {\bibfnamefont {C.}~\bibnamefont {Schubert}},
  \bibinfo {author} {\bibfnamefont {E.~M.}\ \bibnamefont {Rasel}}, \bibinfo
  {author} {\bibfnamefont {N.}~\bibnamefont {Gaaloul}},\ and\ \bibinfo {author}
  {\bibfnamefont {K.}~\bibnamefont {Hammerer}},\ }\bibfield  {title} {\bibinfo
  {title} {{Large-Momentum-Transfer Atom Interferometers with
  $\mathrm{\ensuremath{\mu}}\mathrm{rad}$-Accuracy Using Bragg Diffraction}},\
  }\href {https://doi.org/10.1103/PhysRevLett.131.033602} {\bibfield  {journal}
  {\bibinfo  {journal} {Phys. Rev. Lett.}\ }\textbf {\bibinfo {volume} {131}},\
  \bibinfo {pages} {033602} (\bibinfo {year} {2023})}\BibitemShut {NoStop}%
\bibitem [{\citenamefont {Kirsten-Siem{\ss}}(2023)}]{kirsten:2023:phd}%
  \BibitemOpen
  \bibfield  {author} {\bibinfo {author} {\bibfnamefont {J.-N.}\ \bibnamefont
  {Kirsten-Siem{\ss}}},\ }{\selectlanguage {English}\emph {\bibinfo {title}
  {\textup{Theory of Large-Momentum-Transfer Atom Interferometry in the
  Quasi-Bragg Regime}}}},\ \href {https://doi.org/10.15488/14603} {Ph.D.
  thesis},\ \bibinfo  {school} {Leibniz University Hannover}, \bibinfo
  {address} {Hannover} (\bibinfo {year} {2023})\BibitemShut {NoStop}%
\bibitem [{\citenamefont {Siem\ss{}}\ \emph {et~al.}(2020)\citenamefont
  {Siem\ss{}}, \citenamefont {Fitzek}, \citenamefont {Abend}, \citenamefont
  {Rasel}, \citenamefont {Gaaloul} \emph {et~al.}}]{siemss:2020:analytic}%
  \BibitemOpen
  \bibfield  {author} {\bibinfo {author} {\bibfnamefont {J.-N.}\ \bibnamefont
  {Siem\ss{}}}, \bibinfo {author} {\bibfnamefont {F.}~\bibnamefont {Fitzek}},
  \bibinfo {author} {\bibfnamefont {S.}~\bibnamefont {Abend}}, \bibinfo
  {author} {\bibfnamefont {E.~M.}\ \bibnamefont {Rasel}}, \bibinfo {author}
  {\bibfnamefont {N.}~\bibnamefont {Gaaloul}}, \emph {et~al.},\ }\bibfield
  {title} {\bibinfo {title} {{Analytic theory for Bragg atom interferometry
  based on the adiabatic theorem}},\ }\href
  {https://doi.org/10.1103/PhysRevA.102.033709} {\bibfield  {journal} {\bibinfo
   {journal} {Phys. Rev. A}\ }\textbf {\bibinfo {volume} {102}},\ \bibinfo
  {pages} {033709} (\bibinfo {year} {2020})}\BibitemShut {NoStop}%
\bibitem [{\citenamefont {Giese}(2015)}]{giese:2015:mechanisms}%
  \BibitemOpen
  \bibfield  {author} {\bibinfo {author} {\bibfnamefont {E.}~\bibnamefont
  {Giese}},\ }\bibfield  {title} {\bibinfo {title} {{Mechanisms of matter-wave
  diffraction and their application to interferometers}},\ }\href
  {https://doi.org/https://doi.org/10.1002/prop.201500020} {\bibfield
  {journal} {\bibinfo  {journal} {Fortschr. Phys.}\ }\textbf {\bibinfo {volume}
  {63}},\ \bibinfo {pages} {337} (\bibinfo {year} {2015})}\BibitemShut
  {NoStop}%
\bibitem [{\citenamefont {Shore}(1990)}]{shore:1990:theory}%
  \BibitemOpen
  \bibfield  {author} {\bibinfo {author} {\bibfnamefont {B.}~\bibnamefont
  {Shore}},\ }\href {https://books.google.de/books?id=pJEfAQAAMAAJ} {\emph
  {\bibinfo {title} {{The Theory of Coherent Atomic Excitation, Multilevel
  Atoms and Incoherence}}}},\ A Wiley-Interscience publication\ (\bibinfo
  {publisher} {Wiley},\ \bibinfo {address} {New York},\ \bibinfo {year}
  {1990})\BibitemShut {NoStop}%
\bibitem [{\citenamefont {Lauber}\ \emph {et~al.}(2011)\citenamefont {Lauber},
  \citenamefont {K{\"u}ber}, \citenamefont {Wille},\ and\ \citenamefont
  {Birkl}}]{lauber:2011:pra}%
  \BibitemOpen
  \bibfield  {author} {\bibinfo {author} {\bibfnamefont {T.}~\bibnamefont
  {Lauber}}, \bibinfo {author} {\bibfnamefont {J.}~\bibnamefont {K{\"u}ber}},
  \bibinfo {author} {\bibfnamefont {O.}~\bibnamefont {Wille}},\ and\ \bibinfo
  {author} {\bibfnamefont {G.}~\bibnamefont {Birkl}},\ }\bibfield  {title}
  {\bibinfo {title} {{Optimized Bose-Einstein-condensate production in a dipole
  trap based on a 1070-nm multifrequency laser: Influence of enhanced two-body
  loss on the evaporation process}},\ }\href
  {https://doi.org/10.1103/PhysRevA.84.043641} {\bibfield  {journal} {\bibinfo
  {journal} {Phys. Rev. A}\ }\textbf {\bibinfo {volume} {84}},\ \bibinfo
  {pages} {043641} (\bibinfo {year} {2011})}\BibitemShut {NoStop}%
\bibitem [{\citenamefont {M\"uller}\ \emph
  {et~al.}(2008{\natexlab{b}})\citenamefont {M\"uller}, \citenamefont {Chiow},\
  and\ \citenamefont {Chu}}]{mueller:2008:atom-wave}%
  \BibitemOpen
  \bibfield  {author} {\bibinfo {author} {\bibfnamefont {H.}~\bibnamefont
  {M\"uller}}, \bibinfo {author} {\bibfnamefont {S.-w.}\ \bibnamefont
  {Chiow}},\ and\ \bibinfo {author} {\bibfnamefont {S.}~\bibnamefont {Chu}},\
  }\bibfield  {title} {\bibinfo {title} {{Atom-wave diffraction between the
  Raman-Nath and the Bragg regime: Effective Rabi frequency, losses, and phase
  shifts}},\ }\href {https://doi.org/10.1103/PhysRevA.77.023609} {\bibfield
  {journal} {\bibinfo  {journal} {Phys. Rev. A}\ }\textbf {\bibinfo {volume}
  {77}},\ \bibinfo {pages} {023609} (\bibinfo {year}
  {2008}{\natexlab{b}})}\BibitemShut {NoStop}%
\bibitem [{\citenamefont {Plotkin-Swing}\ \emph {et~al.}(2018)\citenamefont
  {Plotkin-Swing}, \citenamefont {Gochnauer}, \citenamefont {McAlpine},
  \citenamefont {Cooper}, \citenamefont {Jamison} \emph
  {et~al.}}]{plotkin:2018:three-path}%
  \BibitemOpen
  \bibfield  {author} {\bibinfo {author} {\bibfnamefont {B.}~\bibnamefont
  {Plotkin-Swing}}, \bibinfo {author} {\bibfnamefont {D.}~\bibnamefont
  {Gochnauer}}, \bibinfo {author} {\bibfnamefont {K.~E.}\ \bibnamefont
  {McAlpine}}, \bibinfo {author} {\bibfnamefont {E.~S.}\ \bibnamefont
  {Cooper}}, \bibinfo {author} {\bibfnamefont {A.~O.}\ \bibnamefont {Jamison}},
  \emph {et~al.},\ }\bibfield  {title} {\bibinfo {title} {{Three-Path Atom
  Interferometry with Large Momentum Separation}},\ }\href
  {https://doi.org/10.1103/PhysRevLett.121.133201} {\bibfield  {journal}
  {\bibinfo  {journal} {Phys. Rev. Lett.}\ }\textbf {\bibinfo {volume} {121}},\
  \bibinfo {pages} {133201} (\bibinfo {year} {2018})}\BibitemShut {NoStop}%
\bibitem [{Note1()}]{Note1}%
  \BibitemOpen
  \bibinfo {note} {We use $\Omega _\protect \text {R}= 0.42 \times 4P \protect
  \tilde {U}_0(\lambda )/(\hbar \pi w_0^2)$, where $P$, $w_0$, and $\lambda
  =\SI {780.226}{\nano \meter }$ are average power, waist, and wavelength of
  the Bragg beams. $\protect \tilde {U}_0(\lambda )$ is the dipole factor~\cite
  {kuber:2014:dynamics,metcalf:1999:springer,grimm:1999:dipole,steck:2007:quantum}
  for $^{87}$Rb. The factor $0.42$ arises from $f(t)$. This definition has been
  verified by finding good agreement of $\Omega _\protect \text {R}$
  experimentally observed for resonant first-order diffraction and the one
  extracted from simulations.}\BibitemShut {Stop}%
\bibitem [{\citenamefont {Szigeti}\ \emph {et~al.}(2012)\citenamefont
  {Szigeti}, \citenamefont {Debs}, \citenamefont {Hope}, \citenamefont
  {Robins},\ and\ \citenamefont {Close}}]{Szigeti:2012:momentumwidth}%
  \BibitemOpen
  \bibfield  {author} {\bibinfo {author} {\bibfnamefont {S.~S.}\ \bibnamefont
  {Szigeti}}, \bibinfo {author} {\bibfnamefont {J.~E.}\ \bibnamefont {Debs}},
  \bibinfo {author} {\bibfnamefont {J.~J.}\ \bibnamefont {Hope}}, \bibinfo
  {author} {\bibfnamefont {N.~P.}\ \bibnamefont {Robins}},\ and\ \bibinfo
  {author} {\bibfnamefont {J.~D.}\ \bibnamefont {Close}},\ }\bibfield  {title}
  {\bibinfo {title} {{Why momentum width matters for atom interferometry with
  Bragg pulses}},\ }\href {https://doi.org/10.1088/1367-2630/14/2/023009}
  {\bibfield  {journal} {\bibinfo  {journal} {New J. Phys.}\ }\textbf {\bibinfo
  {volume} {14}},\ \bibinfo {pages} {023009} (\bibinfo {year}
  {2012})}\BibitemShut {NoStop}%
\bibitem [{\citenamefont {Auzinger}\ \emph {et~al.}(2017)\citenamefont
  {Auzinger}, \citenamefont {Hofst{\"a}tter}, \citenamefont {Ketcheson},\ and\
  \citenamefont {Koch}}]{auzinger:2017:practical}%
  \BibitemOpen
  \bibfield  {author} {\bibinfo {author} {\bibfnamefont {W.}~\bibnamefont
  {Auzinger}}, \bibinfo {author} {\bibfnamefont {H.}~\bibnamefont
  {Hofst{\"a}tter}}, \bibinfo {author} {\bibfnamefont {D.}~\bibnamefont
  {Ketcheson}},\ and\ \bibinfo {author} {\bibfnamefont {O.}~\bibnamefont
  {Koch}},\ }\bibfield  {title} {\bibinfo {title} {{Practical splitting methods
  for the adaptive integration of nonlinear evolution equations. Part I:
  Construction of optimized schemes and pairs of schemes}},\ }\href
  {https://doi.org/10.1007/s10543-016-0626-9} {\bibfield  {journal} {\bibinfo
  {journal} {BIT Numer. Math.}\ }\textbf {\bibinfo {volume} {57}},\ \bibinfo
  {pages} {55} (\bibinfo {year} {2017})}\BibitemShut {NoStop}%
\bibitem [{\citenamefont {Sidorenkov}\ \emph {et~al.}(2020)\citenamefont
  {Sidorenkov}, \citenamefont {Gautier}, \citenamefont {Altorio}, \citenamefont
  {Geiger},\ and\ \citenamefont {Landragin}}]{sidorenkov:2020:tailoring}%
  \BibitemOpen
  \bibfield  {author} {\bibinfo {author} {\bibfnamefont {L.~A.}\ \bibnamefont
  {Sidorenkov}}, \bibinfo {author} {\bibfnamefont {R.}~\bibnamefont {Gautier}},
  \bibinfo {author} {\bibfnamefont {M.}~\bibnamefont {Altorio}}, \bibinfo
  {author} {\bibfnamefont {R.}~\bibnamefont {Geiger}},\ and\ \bibinfo {author}
  {\bibfnamefont {A.}~\bibnamefont {Landragin}},\ }\bibfield  {title} {\bibinfo
  {title} {{Tailoring Multiloop Atom Interferometers with Adjustable Momentum
  Transfer}},\ }\href {https://doi.org/10.1103/PhysRevLett.125.213201}
  {\bibfield  {journal} {\bibinfo  {journal} {Phys. Rev. Lett.}\ }\textbf
  {\bibinfo {volume} {125}},\ \bibinfo {pages} {213201} (\bibinfo {year}
  {2020})}\BibitemShut {NoStop}%
\bibitem [{\citenamefont {Chih}\ and\ \citenamefont
  {Holland}(2021)}]{chih:2021:reinforcement}%
  \BibitemOpen
  \bibfield  {author} {\bibinfo {author} {\bibfnamefont {L.-Y.}\ \bibnamefont
  {Chih}}\ and\ \bibinfo {author} {\bibfnamefont {M.}~\bibnamefont {Holland}},\
  }\bibfield  {title} {\bibinfo {title} {{Reinforcement-learning-based
  matter-wave interferometer in a shaken optical lattice}},\ }\href
  {https://doi.org/10.1103/PhysRevResearch.3.033279} {\bibfield  {journal}
  {\bibinfo  {journal} {Phys. Rev. Res.}\ }\textbf {\bibinfo {volume} {3}},\
  \bibinfo {pages} {033279} (\bibinfo {year} {2021})}\BibitemShut {NoStop}%
\bibitem [{\citenamefont {Hu}\ \emph {et~al.}(2017)\citenamefont {Hu},
  \citenamefont {Poli}, \citenamefont {Salvi},\ and\ \citenamefont
  {Tino}}]{hu:2017:atominterferometry}%
  \BibitemOpen
  \bibfield  {author} {\bibinfo {author} {\bibfnamefont {L.}~\bibnamefont
  {Hu}}, \bibinfo {author} {\bibfnamefont {N.}~\bibnamefont {Poli}}, \bibinfo
  {author} {\bibfnamefont {L.}~\bibnamefont {Salvi}},\ and\ \bibinfo {author}
  {\bibfnamefont {G.~M.}\ \bibnamefont {Tino}},\ }\bibfield  {title} {\bibinfo
  {title} {{Atom Interferometry with the Sr Optical Clock Transition}},\ }\href
  {https://doi.org/10.1103/PhysRevLett.119.263601} {\bibfield  {journal}
  {\bibinfo  {journal} {Phys. Rev. Lett.}\ }\textbf {\bibinfo {volume} {119}},\
  \bibinfo {pages} {263601} (\bibinfo {year} {2017})}\BibitemShut {NoStop}%
\bibitem [{\citenamefont {K{\"u}ber}(2014)}]{kuber:2014:dynamics}%
  \BibitemOpen
  \bibfield  {author} {\bibinfo {author} {\bibfnamefont {J.}~\bibnamefont
  {K{\"u}ber}},\ }\emph {\bibinfo {title} {\textup{Dynamics of Bose-Einstein
  condensates in novel optical potentials}}},\ \href
  {http://tuprints.ulb.tu-darmstadt.de/4037/} {Ph.D. thesis},\ \bibinfo
  {school} {Technische Universit{\"a}t Darmstadt}, \bibinfo {address}
  {Darmstadt} (\bibinfo {year} {2014})\BibitemShut {NoStop}%
\bibitem [{\citenamefont {Metcalf}\ and\ \citenamefont {Van~der
  Straten}(1999)}]{metcalf:1999:springer}%
  \BibitemOpen
  \bibfield  {author} {\bibinfo {author} {\bibfnamefont {H.~J.}\ \bibnamefont
  {Metcalf}}\ and\ \bibinfo {author} {\bibfnamefont {P.}~\bibnamefont {Van~der
  Straten}},\ }\href@noop {} {\emph {\bibinfo {title} {{Laser cooling and
  trapping}}}}\ (\bibinfo  {publisher} {Springer Science \& Business Media},\
  \bibinfo {address} {New York},\ \bibinfo {year} {1999})\BibitemShut {NoStop}%
\bibitem [{\citenamefont {Grimm}\ \emph {et~al.}(2000)\citenamefont {Grimm},
  \citenamefont {Weidem{\"u}ller},\ and\ \citenamefont
  {Ovchinnikov}}]{grimm:1999:dipole}%
  \BibitemOpen
  \bibfield  {author} {\bibinfo {author} {\bibfnamefont {R.}~\bibnamefont
  {Grimm}}, \bibinfo {author} {\bibfnamefont {M.}~\bibnamefont
  {Weidem{\"u}ller}},\ and\ \bibinfo {author} {\bibfnamefont {Y.~B.}\
  \bibnamefont {Ovchinnikov}},\ }\bibfield  {title} {\bibinfo {title} {{Optical
  dipole traps for neutral atoms}},\ }in\ \href@noop {} {\emph {\bibinfo
  {booktitle} {Advances in atomic, molecular, and optical physics}}},\
  Vol.~\bibinfo {volume} {42}\ (\bibinfo  {publisher} {Elsevier, New York},\
  \bibinfo {year} {2000})\ pp.\ \bibinfo {pages} {95--170}\BibitemShut
  {NoStop}%
\bibitem [{\citenamefont {Steck}(2024)}]{steck:2007:quantum}%
  \BibitemOpen
  \bibfield  {author} {\bibinfo {author} {\bibfnamefont {D.~A.}\ \bibnamefont
  {Steck}},\ }\href
  {https://atomoptics.uoregon.edu/~dsteck/teaching/quantum-optics/} {\bibinfo
  {title} {{Quantum and Atom Optics}}} (\bibinfo {year} {2024}),\ \bibinfo
  {note} {revision 0.16.1}\BibitemShut {NoStop}%
\end{thebibliography}%
\end{document}